\documentclass[aps,pra,twocolumn,floatfix,notitlepage,superscriptaddress,longbibliography]{revtex4-2} 
\usepackage{amsmath}
\usepackage{graphicx}
\usepackage{dcolumn}
\usepackage{xcolor}   
\usepackage{bm}        
\usepackage{amssymb}   
\usepackage{physics}
\usepackage{tikz}
\usepackage{gensymb}
\usepackage{xcolor}
\usepackage{float}
\usepackage{dsfont}
\usepackage[caption = false]{subfig}
\usepackage{mathtools}
 \usepackage{amsthm}
 \usepackage[english]{babel}
\usepackage[normalem]{ulem}
\usepackage[bookmarks,bookmarksopen,bookmarksdepth=2]{hyperref}
\hypersetup{
    colorlinks=true,
    citecolor=blue,
    linkcolor=blue,
    filecolor=magenta,
    urlcolor=blue,
}
 \usepackage{quantikz}
\usepackage[capitalize]{cleveref}
\crefname{section}{Sec.}{Secs.}  
\crefname{appendix}{App.}{Apps.}  
\crefname{Figure}{Fig.}{Figs.} 
\newcommand{\rme}[3]{\langle#1\vert\vert #2 \vert \vert#3\rangle}

\newcommand{\SU}{\mathrm{SU}}
\makeatletter
\DeclareRobustCommand{\element}[1]{\@element#1\@nil}
\def\@element#1#2\@nil{%
  #1%
  \if\relax#2\relax\else\MakeLowercase{#2}\fi}
\makeatother
\let\oldaddcontentsline\addcontentsline

\newcommand{\starttocentries}{\let\addcontentsline\oldaddcontentsline}

\def \addJPMorgan{Global Technology Applied Research, JPMorganChase, New York, NY 10017, USA}
\def \addCQuIC {Center for Quantum Information and Control, University of New Mexico, Albuquerque, NM, USA}

\def \addPandAUNM {Department of Physics and Astronomy, University of New Mexico, Albuquerque, NM, USA}

\def \addLANL {Los Alamos National Laboratory, Los Alamos, NM, USA}

\def \addJPMorgan{Global Technology Applied Research, JPMorganChase, New York, NY 10017, USA}

\begin{document}

\widetext

\title{Coherence Preserving Leakage Detection and Cooling in Alkaline Earth Atoms}

\author{Sivaprasad Omanakuttan}
\email[]{sivaprasad.thattupurackalomanakuttan@jpmchase.com}
\affiliation{\addJPMorgan}\affiliation{\addCQuIC} \affiliation{\addPandAUNM}

\author{Vikas Buchemmavari}
\email[]{bsdvikas@unm.edu}
\affiliation{\addCQuIC} \affiliation{\addPandAUNM}

\author{Michael J. Martin }
\affiliation{\addLANL} \affiliation{\addCQuIC}
\author{Ivan H Deutsch}
\email[]{ideutsch@unm.edu}
\affiliation{\addCQuIC} \affiliation{\addPandAUNM}

\date{\today}
\begin{abstract}

Optically trapped atoms in arrays of optical tweezers have emerged as a powerful platform for quantum information processing given the recent demonstrations of high-fidelity quantum logic gates and on-demand reconfigurable geometry.
Additional errors will occur both in gate operations and atomic transport due to leakage out of the computation space, atomic motional heating, or the complete loss of an atom out of a trap. 
In this work, we address these error channels in a unified manner through laser fluorescence that can detect and cool the atom without disturbing the quantum information encoded therein.  
As only the electrons in the atom couple directly to the laser field, such quantum nondemolition (QND) processes are made possible by encoding quantum information in the nuclear spin of alkaline earth-like atoms and avoiding the effects of the hyperfine interaction which couples it to the electrons.  
By detuning a fluorescence laser off-resonantly from the $\mathrm{^1S_0} \rightarrow \mathrm{^1P_1}$ transition, far compared to the (small) hyperfine spitting, optical pumping between nuclear states falls off rapidly with detuning, scaling as $~1/\Delta^4$. 
In contrast, Rayleigh scattering falls off as $~1/\Delta^2$.
We also consider a resonant leakage detection protocol off the $^1\mathrm{P}_1$ line.
This is achieved by disabling the hyperfine coupling via a strong AC stark effect and canceling the residual lightshifts via dressing. 
The same scheme can be modified to recool the atoms towards the vibrational ground state for the quantum information encoded in the ground state of alkaline earth atoms while preserving the coherence.  
These advances could significantly improve the prospect of neutral atoms for fault-tolerant quantum computation.

\end{abstract}
\maketitle
\section{Introduction}


Quantum information processing (QIP) with neutral atoms has emerged as a promising platform in recent times due to their unique capabilities such as the potential to scale to thousands of high-quality qubits~\cite{Bluvstein_Lukin_2023_logical,Pritchard_single_qubit_4nines_PRL_2023, pichard2024rearrangementsingleatoms2000site, manetsch2024tweezerarray6100highly,norcia2024iterative}, the ability to rearrange atoms to arbitrary connectivity~\cite{Barredo_Browaeys_Atom_assembly, Endres_Lukin_Atom_assembly,Lukin_Nature_2022,Bluvstein_Lukin_2023_logical}, and the demonstration of many parallel high-fidelity entangling gates using Rydberg interactions~\cite{Evered_Lukin_2023_High_fidelity,peper2024spectroscopymodeling171ybrydberg,finkelstein2024universalquantumoperationsancillabased,cao2024multiqubitgatesschrodingercat}.
 These features have enabled neutral atoms to become a leading platform for quantum metrology~\cite{Schine_Kaufman_Bell_state_Martin,Kaubruegger_Zoller_metrology,Bornet_Browaeys_2023_Scalable_sqeezing,Hines_Schleier-Smith_2023_Squeezing_Rydberg_dressing, Eckner_Kaufman_2023_Squeezing_clock}, quantum simulation~\cite{Scholl_Browaeys_Nature_2021_Q_simulation,Ebadi_Lukin_Nature_2021_256_atom,Bornet_Browaeys_2023_Scalable_sqeezing,Browaeys_many_body_review_2020_Nature,Bornet_Browaeys_2023_Scalable_sqeezing}, and {quantum computation~\cite{Evered_Lukin_2023_High_fidelity,Bluvstein_Lukin_2023_logical,Saffman_Nature_2022}}.
 Achieving fault tolerance using quantum error correction (QEC) techniques is an ultimate goal for every physical quantum system. 
 To this end, many advances are being made both experimentally (such as mid-circuit measurements~\cite{graham2023midcircuit,lis2023mid,mid_circuit_atom_computing}, erasure conversion~\cite{Ma_Thompson_Yb_erasure_gate,Scholl_Endres_2023_erasure_simulation}, erasure cooling~\cite{scholl2023erasure_cooling}) and theoretically (such as fault-tolerance frameworks~\cite{Cong_Lukin_QEC_Rydberg_PRX_2022,Wu_Puri_Thompson_2022_Nature_erasure,Omanakuttan_spincats_PRXQ_2024,PhysRevA.108.022428}, QEC codes that leverage hardware capabilities~\cite{Wu_Puri_Thompson_2022_Nature_erasure,Sahay_biased_erasure_PRXQ_2023}, and efficient gate designs).
 
 Early work in neutral atom quantum computing was done mainly using alkali metal atoms like cesium~\cite{Saffman_Isenhower_CNOT_PRL_2010} and rubidium~\cite{Levine_Pichler_gate}. State-of-the-art alkali atom systems continue to make great strides, including the landmark experiment demonstrating logical operations on 48 logical qubits~\cite{Bluvstein_Lukin_2023_logical}. Paralleling the development of optical atomic clocks, alkaline-earth (like) atoms have also emerged as promising contenders such as $^{88}\mathrm{Sr}$~\cite{Madjarov_Endres_2020_Sr,Scholl_Endres_2023_erasure_simulation,Schine_Kaufman_Bell_state_Martin,Eckner_Kaufman_2023_Squeezing_clock}, $^{171}\mathrm{Yb}$~\cite{Lis_Senoo_Kaufman_Yb_mid_circuit_2023,Ma_Thompson_Yb_erasure_gate,Covey_Yb,mid_circuit_atom_computing}, $^{87}\mathrm{Sr}$~\cite{Atom_computing_2022, omanakuttan2021quantum}. 
This wide breadth of architectural developments has led to a rich landscape of possible techniques with various trade-offs. While substantial progress has been made, further advances are necessary on the path to fault tolerance.  This will involve both reducing sources of error to improve the fidelity of operations and improving codes to be more tolerant of some errors, thereby substantially improving thresholds and reducing the required resources.

A dominant source of imperfection in the neutral atom platform is the "leakage errors," which take the state outside the computational subspace. 
There are multiple sources of such errors including atoms escaping the weak trapping potential, population getting trapped in a metastable state such as Rydberg states during gates implementations, collisions with background gas, etc. 
Leakage errors are detrimental to the prospects of fault tolerance as they are not the standard Pauli errors and require resource-expensive leakage reduction units (LRU) to tackle these errors~\cite{Suchara_2015_leakage,chow2024_erasure}.
An alternative to this is to convert leakage errors to ``erasure errors'', where we know the location of the leakage errors which can then be corrected very efficiently.  

Of particular interest are the alkaline-earth (like) atoms, strontium, and ytterbium, also employed in optical clocks.
 For example, by encoding quantum information in the nuclear spin of $^{171}$Yb in the metastable clock state $\mathrm{^3P_0}$, errors could be dominated by leakage that can be detected and erased~\cite{Wu_Puri_Thompson_2022_Nature_erasure}. 
 Leakage occurs during two-qubit gates when atoms are excited to high-lying Rydberg states with a finite lifetime.  
 Two major leakage channels are black-body induced transitions to other metastable Rydberg states and decay to the ground $\mathrm{^1S_0}$.  
 The latter can be detected by simple fluorescence without disturbing quantum information encoded in the nuclear spin in metastable $\mathrm{^3P_0}$ state.  
 Such erasure-dominant error correction schemes can greatly improve the thresholds for fault tolerance~\cite{Wu_Puri_Thompson_2022_Nature_erasure,Sahay_biased_erasure_PRXQ_2023}.
 However, this protocol does not account for ``atom loss'', which significantly hinders the potential of using neutral atoms as viable candidates for fault-tolerant quantum computers.
 Further, metastable encoding schemes suffer from optical pumping that increases the overall error rate that must be converted to erasures. Also, for architectures such as in~\cite{Bluvstein_Lukin_2023_logical,Lukin_Nature_2022}, one requires atomic transport which will heat the atomic motion. In trapped atomic ions, sympathetic cooling in shared vibrational motion with a distinct refrigerant atomic species is used to recool atoms after transport~\cite{sympathetic_cooling_Wineland,sympathetic_cooling_Wineland_Monroe}. 
Such a direct mechanism is not straightforward for neutral atoms (see related work~\cite{GorshkovRydbergCooling}).

Our goal, thus, is twofold. 
We need to convert the leakage errors (including atom loss) to erasure errors as well as to recool the atoms towards the vibrational ground state for the quantum information encoded in the ground $\mathrm{^1S_0}$ state.
We must achieve this without disturbing the quantum information stored therein; this constitutes a  Quantum Non-Demolition (QND) process.
In the ground $\mathrm{^1S_0}$ state of alkaline-earth (like) atoms, quantum information can be encoded in the nuclear spin for fermionic isotopes.
There is no hyperfine interaction in this manifold, so the quantum information encoded therein is highly isolated from the environment. 
The laser light that is used to image and cool atoms directly couples only to the electrons, and indirectly to the nuclei, only through the hyperfine interaction.  
By scattering photons from atoms in a way that avoids hyperfine coupling, we can achieve both of these goals in a QND manner. 
Whereas previous work considered decoupling electron angular momentum and nuclear spin through the use of a large magnetic field~\cite{Reichenbach_Deutsch_cooling_PRL_2007}, we consider a more flexible approach here based on large AC-Stark shifts as proposed by Shi~\cite{Shi_cooling_PRA_2023}. 
We will show how one can retain coherence across all magnetic sublevels in the nuclear spin with $I\ge 1/2$, thus compatible with new protocols that employ multiple levels for qudit gates~\cite{omanakuttan2022qudit,omanakuttan2021quantum} and error correction~\cite{omanakuttan2023multispin, Omanakuttan_spincats_PRXQ_2024}.

The remainder of this paper is organized as follows.
After establishing the necessary background in \cref{sec:background}, in \cref{sec:qnd_leakage_detection} we propose and analyze a protocol that converts leakage errors to erasure errors for alkaline-earth (like) atoms through Rayleigh scattering of photons.
In \cref{sec:qnd_resonance_fluorescence_and_cooling}, we describe the general principles of resonance fluorescence while preserving coherence.
In \cref{sec:qnd_cooling} we augment the protocol to include resolved sideband cooling of the atoms while preserving nuclear spin coherence.
This scheme generalizes the proposed schemes in \cite{Reichenbach_Deutsch_cooling_PRL_2007, Shi_cooling_PRA_2023}.
We {summarize and conclude}  in \cref{sec:conclusions_and_future_work}.

\section{Background}
\label{sec:background}

\begin{figure}[!ht]
    \centering
    \includegraphics[width=0.9\columnwidth]{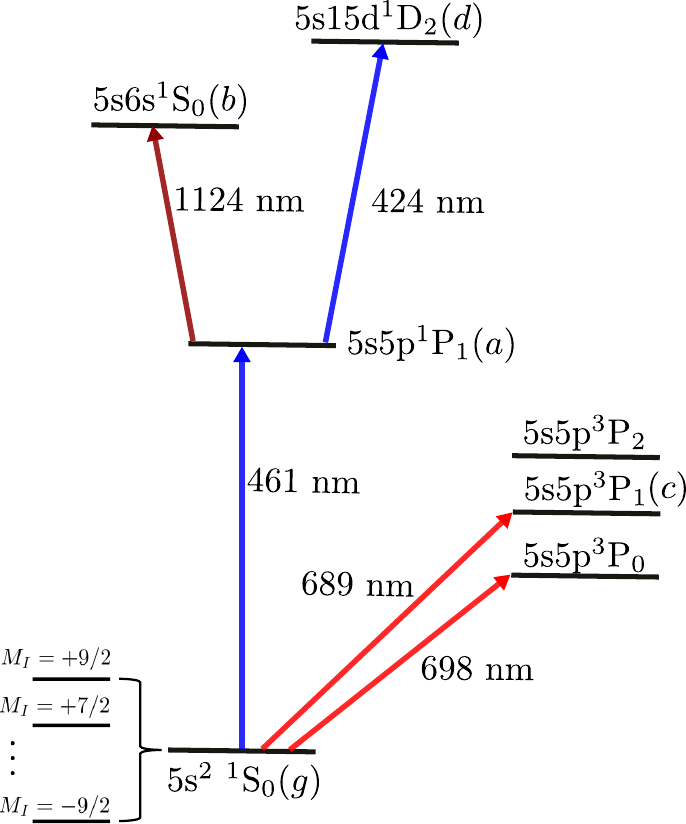}
    \caption{ Atomic level structure of $^{87}$Sr.
    We encode quantum information in the $I=9/2$ nuclear spin in the electronic ground state, 5s$^2$ $^1\mathrm{S}_0$.
      One can encode a qudit with dimension $ d\leq 10$ in the 10 nuclear Zeeman sublevels. The $^3\mathrm{P}_0$ clock manifold is long-lived and has no hyperfine coupling due to electronic spin $J=0$.
      The $^3\mathrm{P}_1$ has moderate linewidth and strong hyperfine coupling and can be used to induce quadratic lightshifts in the ground manifold~\cite{omanakuttan2021quantum}. 
      Population in the $5\mathrm{s}^2~^1\mathrm{S}_0$ can be detected by coupling to the broad linewidth $^1\mathrm{P}_1$ manifold and detecting scattered photons. 
      The $^1\mathrm{P}_1$ manifold has weaker hyperfine coupling, which causes decoherence in the nuclear spin encoded information during fluorescence. 
      The hyperfine interaction can be ignored for large detuning but the ground manifold experiences a residual quadratic light shift that can be canceled. 
      Alternatively, the hyperfine interaction can be suppressed by dressing with the 5s6s$^1\mathrm{S}_0$ and 5s15d$^1\mathrm{D}_2$.
      }
    \label{fig:basic_outline}
\end{figure}

In this work, we focus on encoding quantum information in the nuclear spin $I$ in the  $^1\mathrm{S}_0$ electronic ground state of fermionic alkaline-earth (like) elements.  The nuclear spin is highly isolated from the environment and thus serves as a robust memory for quantum information \cite{barnes2021assembly,PhysRevLett.101.170504,daley2011quantum}.
Specifically, we focus on $^{87}\mathrm{Sr}$, which has nuclear spin $I=9/2$ in the main text, and address Yb in \cref{sec:qnd_leakage_detection_yb,sec:qnd_cooling_yb}.

 The relevant level diagram is given in \cref{fig:basic_outline} (for a comprehensive review see, e.g.,~\cite{katori2002spectroscopy,martin2013quantum}). 
 The ground state of the atom is the spin-singlet 5s$^2$ $^1\mathrm{S}_0$ with zero electronic spins $J$. Hence the total angular momentum $F$ of the ground state only has a contribution from nuclear spin, $F=I=9/2$.
This results in a $d = 2I+1=10$ ten-dimensional ground manifold, comprising a 10-dimensional qudit. The qudit states are labeled as  $\ket{0}=\ket{M_I=9/2}, \ket{1}=\ket{M_I=7/2},\cdots ,\ket{9}=\ket{M_I=-9/2}$.
In previous work \cite{omanakuttan2021quantum,omanakuttan2022qudit} we {showed} how one can implement arbitrary $\SU(10)$ maps using quantum optimal control and {implement entangling gates using Rydberg interactions on these qudits}. 
 These multiple levels can also encode a logical qubit in a spin-cat error-correcting code, significantly reducing the stringent requirements for fault-tolerant quantum computation \cite{Omanakuttan_spincats_PRXQ_2024}. 
Encoding quantum information in the ground state of the $^{87}$Sr has potential application in quantum simulation~\cite{zache2023fermion,zhang2014spectroscopic}. 

To good approximation, the excited states can exist in either a spin-singlet or triplet configuration.
 As singlet-triplet transitions are forbidden, the intercombination lines, e.g. $\mathrm{5s^2\text{ }^1S_0}\rightarrow \mathrm{5s5p\text{ }^3P}_J$, are very narrow linewidth.  
 The $\mathrm{5s5p\text{ }^3P_1}$ state has a large hyperfine splitting ($\sim 2.5$ GHz), well resolved compared to the $7.5$ kHz linewidth.  The $\mathrm{5s5p\text{ }^3P_0}$ ``clock state," with $J\approx 0$, has no hyperfine splitting and is highly forbidden to decay to the ground state.  
 In contrast, the decay of the $\mathrm{5s5p\text{ }^1P_1}$ state is strongly dipole allowed, yielding a very short lifetime. 
 The hyperfine splitting in this state is small ($\sim 60$ MHz) and not well resolved when compared to the 32 MHz linewidth.  
 In addition, we can optically couple the singlet $\mathrm{5s5p\text{ }^1P_1}$ state to excited 5s6s$\mathrm{^1S_0}$ and 5s15d$\mathrm{^1D_2}$  
 \footnote{We generally identify the 5s$^2$ $^1\mathrm{S}_0$ and 5s5p$^1\mathrm{P}_1$ states simply as $^1\mathrm{S}_0$ and $^1\mathrm{P}_1$ respectively. The excited state 5s6s$^1\mathrm{S}_0$ is denoted fully or appended with the word appended with the word \textit{excited} to avoid confusion  } states to manipulate the desired interactions.  
 We will leverage all these features for different applications in QND leakage detection and cooling, discussed below.


\section{QND leakage detection via off-resonance Rayleigh scattering}
\label{sec:qnd_leakage_detection}

 We seek to detect leakage out of the electronic ground state.  
 This could occur due to the population remaining in a metastable excited state, or atoms lost from the trap altogether. 
 These are prevalent error channels in current two-qubit entangling gate protocols, as the population in the excited Rydberg state can decay to other nearby Rydberg states, stimulated by black-body radiation.
 As atoms remaining in a Rydberg state typically will be repelled by trapping fields, the atoms will be lost. 
These leakage errors will impose more stringent requirements for fault tolerance~\cite{Suchara_2015_leakage} than Pauli errors.
However, if we can convert the leakage errors to erasure errors, where we know the location of the qubit/qudit in which a leakage has occurred, one can correct these errors more efficiently than the Pauli errors~\cite{Wu_Puri_Thompson_2022_Nature_erasure}.

We thus seek to detect whether the atom is the $\mathrm{^1S_0}$ ground state without disturbing the quantum information stored in the nuclear spin. 
We can achieve this through fluorescence if the photons do not carry any ``which way" information about the nuclear spin state.  
This information may reside in the frequency or polarization of the photon, or the spin-dependent rate at which photons are scattered.   
We consider first far-off resonance photon scattering on the strong  $\mathrm{5s^2\text{ }^1S_0} \rightarrow \mathrm{5s5p\text{ }^1P_1}$ transition.   
In this case, for low saturation of the atomic transition, photons are elastically scattered, with negligible frequency shift in fluorescence (explained in detail below). The remaining which-way information could reside in the photon polarization, which occurs for photon scattering that leads to optical pumping between the nuclear spin magnetic sublevels (see~\cref{subsec:polarization_decoherence}). 
We seek to suppress such optical pumping and ensure only elastic Rayleigh scattering, and to have the rate of scattering be independent of the magnetic sublevel.

Optical pumping between nuclear spin states occurs due to its hyperfine interaction with the electron angular momentum, which exists only in the excited state manifold.  
By detuning far compared the excited-state hyperfine splitting, the hyperfine coupling rapidly goes to zero. 
We show below that in this regime, the rate of optical pumping rapidly falls off with detuning as $\Delta_L^{-4}$ while the Rayleigh scattering rate falls off as  $\Delta_L^{-2}$ so that one can collect sufficient fluorescence photons with little damage to the nuclear spin state. 
The off-resonance coupling will lead also to coherent dynamics of the spin state due to the induced AC-Stark shift, but this can be reversed with appropriate additional laser interaction.

\begin{figure*}[!ht]
    \centering    
    \subfloat[]{\includegraphics[width =0.95\columnwidth]{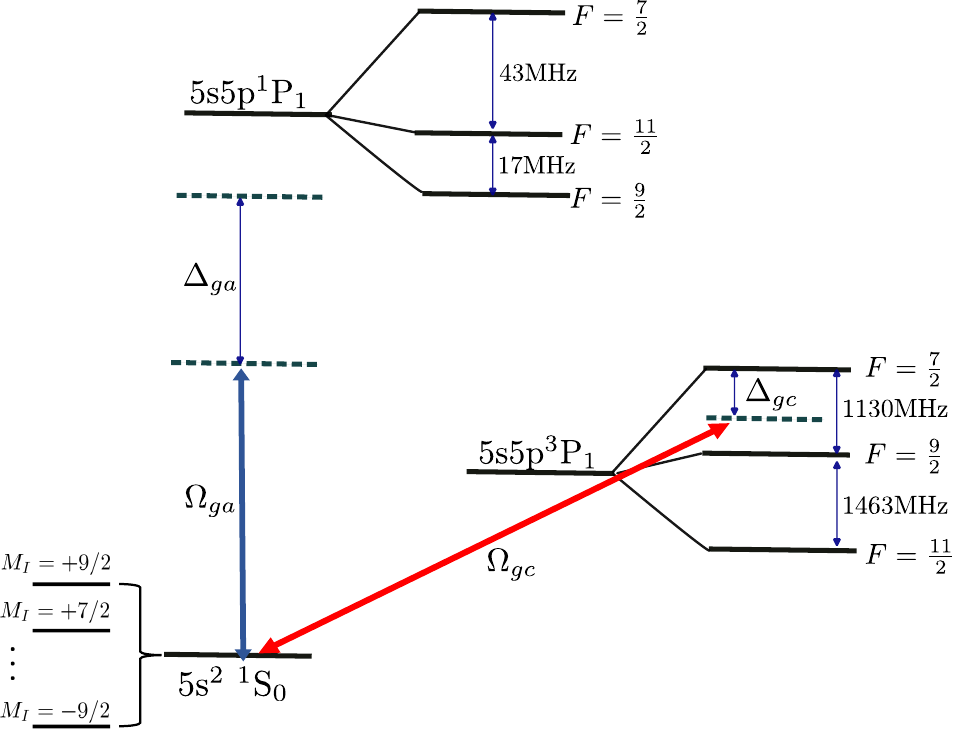} \label{fig:qnd_leakage_a}}\hspace*{1 em}
    \subfloat[]{\includegraphics[width =1.05\columnwidth]{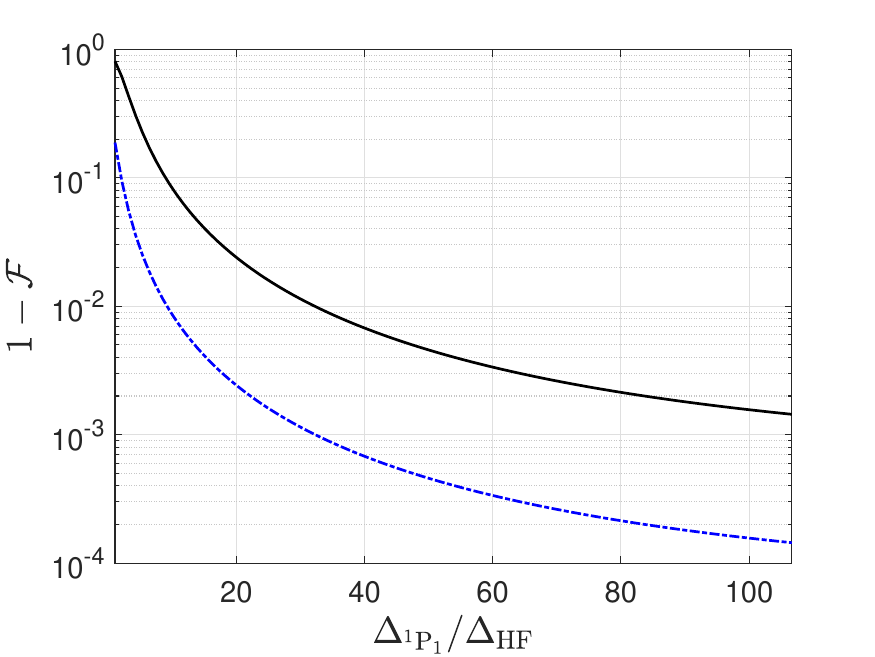}  \label{fig:qnd_leakage_b}}   
    \caption{ QND detection of the $^1\mathrm{S}_0$ ground-state population via far-off resonance Rayleigh scattering. 
    (a) Level diagram of $^{87}$Sr with the requisite laser couplings.
    We use the far-detuned coupling to the singlet $^1\mathrm{P}_1$ manifold, which suppresses optical pumping and decoherence. 
    The hyperfine coupling in $^1\mathrm{P}_1$ is weak and hence $\Delta_{gc}\gg A,Q$ is easily achievable.
    The residual tensor light shift is canceled by dressing with the $^3\mathrm{P}_1$ manifold, with the optimal detuning $\Delta_{gc}$~\cite{omanakuttan2021quantum} and the appropriate Rabi frequency $\Omega_{gc}$.
    (b) Simulated infidelity of the equal superposition state (\cref{eq:initial_state_leakage}) as a function of detection detuning $\Delta_{ga}$. 
    The solid(dashed) line depicts the infidelity for an illumination period of enough for 100(10) scattered photons. Lower infidelity indicates a more effective QND scheme, which can be achieved with a larger detuning but comes at the cost of larger detection time ($\propto \Delta_{ga}^2$). The jump operators corresponding to spontaneous emission from the $^3\mathrm{P}_1$ manifold are also included in these simulations.  }
    \label{fig:qnd_leakage}
\end{figure*}

The basic scheme is shown in \cref{fig:qnd_leakage_a}.  Let the ground state manifold be donated $\mathrm{5s^2\text{ }^1S_0} \equiv g$ and excited state $ \mathrm{5s5p\text{ }^1P_1} \equiv a$. 
We consider a laser field with Rabi frequency $\Omega_{ga}$, off-resonance, with detuning $\Delta_{ga}$ large compared to the excited state hyperfine splitting. 
An additional field drives the $ g \rightarrow \mathrm{5s5p\text{ }^3P_1}\equiv d$ intercombination transition, with a detuning near the highly resolved hyperfine levels, but far from resonance.
The strongly allowed singlet transition provides the fluorescence photons for leakage detection, and the weak triplet transition acts to cancel the effects of light shift.

Consider a state in the computational subspace, 
\begin{equation}
\ket{\psi}=\sum_{i=-\frac{9}{2}}^{\frac{9}{2}}\alpha_i\ket{g, \text{ }M_i},
\end{equation}
where  $\sum_i \lvert \alpha_i \rvert^2=1$. 
The goal of the QND leakage detection scheme is to measure whether or not the atom is in this computational subspace without destroying the quantum information in the state. We quantify the performance of the protocol by calculating the fidelity between the evolved state $\rho$ and the initial state, $\mathcal{F}=\langle \psi |\rho|\psi\rangle$.

To calculate this fidelity, consider first the dynamics induce by coupling the ground state to the $\mathrm{^1P_1} \equiv a$  manifold.   
After adiabatic elimination of the excited state, the ground state $\rho$  evolves according to the  Lindblad Master equation~\cite{deutsch2010quantum},
\begin{equation}
\begin{aligned}
\frac{d\rho}{dt}&=-i\left[H_{\mathrm{LS}},\rho\right]\\
&+ \Gamma_{a}\sum_q W_q^{a}\rho \left(W_q^{a}\right)^{\dagger}-\frac{\Gamma_{a}}{2} \{\left(W_q^{a}\right)^{\dagger}W_q^{a},\rho\}.
\end{aligned}
\end{equation}
Here
\begin{equation}
H_{\mathrm{LS}}= \sum_{F'}\frac{\Omega_{ga}^2}{4\Delta_{g,aF'}} C^{(2)}_{J'=1,F',F} F_z^2 ,
\end{equation}
where, $H_{\mathrm{LS}}$ is the tensor light shift Hamiltonian.  The coefficient $C^{(2)}$ characterizes the rank-2 irreducible polarizability as given in~\cite{deutsch2010quantum} and $F_z$ is a nuclear spin angular momentum operator.  Optical pumping is described by the jump operators,
\begin{widetext}
\begin{equation}
\begin{aligned}
W_q^{a}&=\sum_{F'} \frac{\Omega_{ga}/2}{\Delta_{g,aF'}+i\Gamma_{a}/2}\left[ C^{(0)}_{1FF'}\left(\bm{e}_q^{*}.\vec{\epsilon}_L\right) \mathds{1}+iC^{(1)}_{1FF'}\left(\bm{e}_q^{*}\cross\vec{\epsilon}_L\right).\bm{F} +C^{(2)}_{1,F,F'}\left(\frac{\left(\bm{e}_q^{*}.\bm{F}\right)\left(\vec{\epsilon}_L.\bm{F}\right)+\mathrm{h.c}}{2}-\frac{1}{3}\lvert \bm{e}_q^{*}.\vec{\epsilon}_L\rvert^2\bm{F}^2 \right) \right],
\end{aligned}
\end{equation}
\end{widetext}
where $\Gamma_{a}$ is the rate of spontaneous decay of the excited state $a$, $\vec{\epsilon}_L$ is the polarization of the laser, and $q={-1,0,1}$ represents the polarization of the scattered light in the spherical basis.  For $\pi$-polarized light,  defining $\Delta_{g,aF'}=\Delta_{ga}+\delta_{aF'}$, and working in a far-off resonance regime where the detuning is much larger than the hyperfine splitting ($\Delta_{ga} \gg \delta_{aF'}$), then to lowest order,
\begin{equation}
    H_{\mathrm{LS}} \approx \frac{\Omega_{ga}^2}{4 \Delta_{ga}} \mathds{1}-\frac{\Omega_{ga}^2}{4 \Delta_{ga}^2}\beta^{(2)}F_z^2, 
\end{equation}
where we have used the fact that for the coupling between the electronic angular momentum's $J=0$ and $J'=1$,
\begin{equation}
\begin{aligned}
\sum_{F'} C^{(2)}_{1FF'}=0,\\
\end{aligned}
\label{eq:sum_of_coefficients}
\end{equation}
and for convenience of notation, we have defined,
\begin{equation}
\begin{aligned}
\beta^{(i)}=\sum_{F'}C^{(i)}_{1F'F} \delta_{aF'}.
\end{aligned}
\end{equation}
Retaining terms up to $\mathcal{O}(1/\Delta_{ga}^2)$ , the general form of the jump operators is
\begin{widetext}
\begin{equation}
\begin{aligned}
W_0&\approx \frac{\Omega_{ga}}{2\Delta_{ga}}\mathds{1}+\frac{\Omega_{ga}}{2\Delta_{ga}^2}\left(\left(\beta^{(0)}+i\gamma^{(0)} \right)\mathds{1}+\left(\beta^{(2)}+i\gamma^{(2)} \right)F_z^2 \right) ,\\
W_{\pm}&\approx \frac{\Omega_{ga}}{2\Delta_{ga}^2} \left(i\left(\beta^{(1)}+i\gamma^{(1)} \right)F_{\mp} + \left(\beta^{(2)}+i\gamma^{(2)} \right)\left[\frac{F_zF_{\mp}+F_{\mp}F_z}{2}\right]\right),\\
\end{aligned}
\end{equation}
\end{widetext}

where,
\begin{equation}
\gamma^{(i)}=\frac{\Gamma}{2}\sum_{F'} C^{(i)}_{1F'F}.
\label{eq:gamma_equation}
\end{equation} 
We highlight a few facts. 
The dominant effect in the jump operators is the scalar term in $W_0$, which corresponds to Rayleigh scattering, does not involve any couplings between the magnetic sublevels.  
The remaining correction terms in the jump operators (arising vector and tensor polarizability) are due to residual hyperfine coupling, but these terms fall off rapidly for large detuning.
Unlike alkali elements where $W_\pm \propto 1/\Delta_{ga}$~\cite{deutsch2010quantum}, $W_{\pm}$  goes as $1/\Delta_{ga}^2$  for the alkaline-earth (like) elements in the far-off resonance regime. 
In addition, for this transition, $\gamma^{(2)}=0$.  This arises from the fact that there is no coupling of the electronic and nuclear degrees of freedom in the ground state. 
A similar study of off-resonance Rayleigh scattering was considered by~\cite{Gorshkov2009} for alkaline-earth (like) atoms in a different regime.
The dynamics of the nuclear spin thus arise only from hyperfine interaction in the excited states. 
In the presence of the far-off resonance driving on the $g\rightarrow a$ transition, the spin evolves according to
\begin{equation}
    \frac{d\rho}{dt} = -i [H^{(2)}_{LS},\rho] +\Gamma_{a}\mathcal{O}\left(\frac{\Delta_{\mathrm{HF}}^4}{\Delta_{ga}^4}  \right),
\end{equation}
where
\begin{equation}
H^{(2)}_{LS}=\frac{\Omega_{ga}^2}{\Delta_{ga}^2} \beta^{(2)} F_z^2
\end{equation}
is the residual tensor light shift.  Rayleigh scattering and the Hamiltonian evolution due to the light shit scale as $\Delta_{ga}^{-2}$, while optical pumping falls off as  $\Delta_{ga}^{-4}$.

To complete the protocol we must cancel the residual tensor light shift without decohering the nuclear spin. To achieve this we can employ a second laser field and use the large tensor light-shift term when coupling to the $\mathrm{^3P_1}\equiv c$ manifold, as studied in detail in~\cite{omanakuttan2021quantum}. 
We choose a detuning of $\Delta_{gc}/2\pi=635$ MHz, about halfway between the hyperfine manifolds $F'=7/2$ and $F'=9/2$, where the ratio of the tensor light shift to decay rate is minimized. The Rabi frequency $\Omega_{gc}$ is chosen such that this light-shift cancels the light-shift due to the QND  laser field. 

To quantify the performance of the scheme in \cref{fig:qnd_leakage_b}, we consider the  initial state to be an equal superposition of all magnetic sublevels in the $\mathrm{^1S_0}$ ground state 
\begin{equation}
\ket{\psi}=\frac{1}{\sqrt{10}}\sum_{M_I=-\frac{9}{2}}^{\frac{9}{2}} \ket{g, M_I},
\label{eq:initial_state_leakage}
\end{equation}
After the cancellation of the light shift, the system evolves according to the master equation,
\begin{equation}
\begin{aligned}
    \frac{d\rho}{dt}&=\Gamma_{a}\sum_q W_q^{a}\rho \left(W_q^{a}\right)^{\dagger}-\frac{\Gamma_{a}}{2} \{\left(W_q^{a}\right)^{\dagger}W_q^{a},\rho\}\\
    &+\Gamma_{c}\sum_q W_q^{c}\rho \left(W_q^{c}\right)^{\dagger}-\frac{\Gamma_{c}}{2} \{\left(W_q^{c}\right)^{\dagger}W_q^{c},\rho\},
\end{aligned}
    \end{equation}
where $W_q^{a/c}$ are the jump operators for the singlet and triplet $\mathrm{P_1}$ states respectively.
In \cref{fig:qnd_leakage_b}, we show the fidelity of the state with respect to the initial state after the time required for Rayleigh scattering $10$ and $100$ photons, off resonantly from the $\mathrm{^1P_1}$ state. We note that the resulting infidelity is proportional to the number of photons scattered. With 100 total photons scattered, a detection efficiency of 10-20\% could be enough to determine the presence of an atom.
From the simulations shown in \cref{fig:qnd_leakage_b}, one can infer that as we increase $\Delta_{ga}$  one can recover the ideal fidelity, and for sufficiently large detunings gives us a QND leakage detection scheme. 
The measurement time increases as $\Delta_{ga}^2$ for a constant number of photons scattered, setting up a tradeoff between illumination time and fidelity loss.

By contrast, the erasure conversion protocol in \cite{Wu_Puri_Thompson_2022_Nature_erasure} considers encoding a qubit in a metastable clock manifold for correcting leakage errors from the Rydberg interaction that is used to generate entangling interactions.
In that case, a detection protocol was designed for each kind of major leakage pathway, such that the measurement results in a non-zero signal (for example, detection of scattered photons) when \textit{that specific leakage occurs}.
Hence, each round of erasure conversion requires different types of detection. 
Atom loss cannot be currently detected in this manner. This also means that in the absence of leakage, the atom is not impacted beyond idling errors. 
Alternatively, we consider encoding information in the ground manifold.
The protocols we propose work to detect all the sources of leakage errors from a single measurement. A nonzero signal is generated when \textit{there is no leakage}. This requires only one round of erasure conversion to detect any kind of leakage, including atom loss.  This comes at the cost that every scattered photon reduces the fidelity of the state by a constant amount.

\section{QND Resonance Fluorescence}
\label{sec:qnd_resonance_fluorescence_and_cooling}

In principle, one can increase the speed of the QND leakage detection and also remove the effect of the tensor light shift by tuning the fluorescence laser on-resonance to the $g\rightarrow a$ transition, in the presence of an appropriate perturbation to suppress the hyperfine effects.  For the $\mathrm{5s5p\;^1P_1}$ state, the  hyperfine interaction is,
\begin{equation}
H_{\mathrm{hf}}=A (\mathbf{I}\cdot\mathbf{J})+Q \frac{3 (\mathbf{I}\cdot\mathbf{J})^2+3 \mathbf{I}\cdot\mathbf{J}/2-I(I+1)J(J+1)}{2I J(2I-1)(2J-1) },
\label{eq:Hyperfine_Hamiltonian}
\end{equation}
 with $A/2\pi=-3.4~\mathrm{MHz}$ and $Q/2\pi=39~\mathrm{MHz}$. 
 Both the magnetic dipolar and quadrupolar terms can lead to decoherence in the ground nuclear spin state due to the ``which-way information'' contained in the polarization and frequency of the spontaneously emitted photons.  The former leads to optical pumping between magnetic sublevels and the latter can lead to loss of coherence in Zeeman state superpositions.
 Below, we show how to suppress these effects with additional laser fields.

\subsection{Removing polarization dependent decoherence}\label{subsec:polarization_decoherence}
In the excited state, $F$ and $M_F$ are the good quantum numbers.
Hence when we resonantly excite to the $\mathrm{^{1}P_1}$ manifold, a single $M_I$ state in the ground state couples to multiple $M_I$ values with different $M_J$ values in the excited state. 
This presence of different $M_J$ values allows the spontaneously emitted photon to have different polarizations, leading to optical pumping \cite{deutsch2010quantum} and nuclear spin decoherence.
The key is then to ensure that when we transfer the population to the $\mathrm{5s5p^{1}P_1}$ manifold, only a single $M_J$ state is allowed.


\begin{figure}
    \centering
    \includegraphics[width=1\columnwidth]{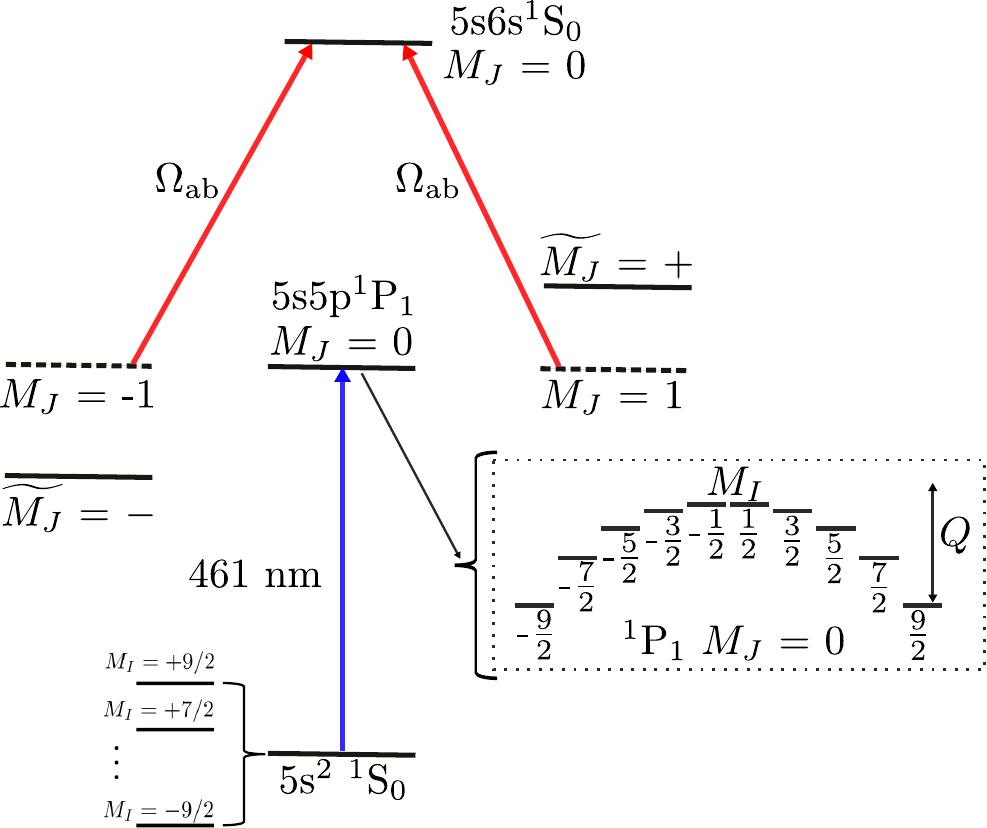}
    \caption{ Effects of strong AC stark effect between 5s5p$^1\mathrm{P}_1$ and 5s6s$^1\mathrm{S}_0$.
    A $\sigma_\pm$ polarized light resonantly couples the $^1\mathrm{P}_1, M_J=\pm1$ manifolds to the 5s6s$^1\mathrm{S}_0$ manifold.     
    This imparts strong light shifts ($\pm\Omega_{ab}/2$) on the $M_J=\pm1$ manifolds. 
    When $\Omega_{ab}\gg A,Q$ the hyperfine coupling is broken and the product basis $\ket{J,M_J}\otimes\ket{I,M_I}$ becomes the good basis. The residual quadrupolar hyperfine interaction induces a quadratic shift, $\approx Q'M_I^2$, on the nuclear spin sublevels.  }
        \label{fig:S_dressing}
\end{figure}
\begin{figure*}[!ht]
    \centering    
     \includegraphics[width=2\columnwidth]{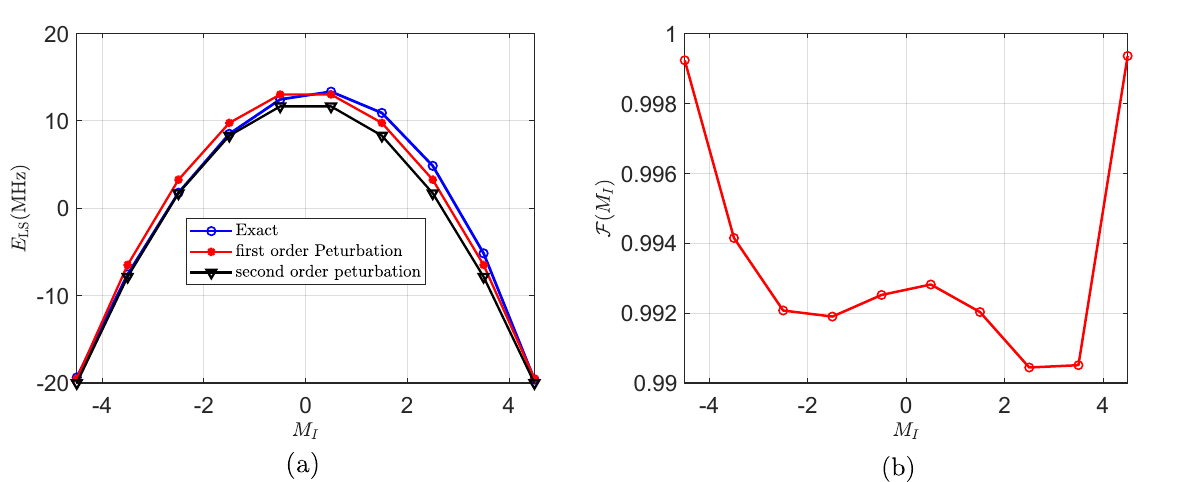} 
    \caption{ 
    Numerical analysis of the effects of strong AC stark effect via coupling between 5s5p$^1\mathrm{P}_1$ and 5s6s$^1\mathrm{S}_0$ shown in \cref{fig:S_dressing} for $\Omega_{{ab}}/2\Pi =1000$ MHz.
    (a) The eigenvalues of the eigenstates $|M_J=0, M_I\rangle$,  the states of interest, and compared with the perturbation theory analysis.
    (b) The overlap of the eigenvectors with the states $\ket{M_J=0,M_I}$, the near $1$ overlap, indicating that the $M_J=0$ state is isolated via the coupling between $\mathrm{5s5p^1P_1}$ and $\mathrm{5s6s^1S_0}$.
    }
    \label{fig:cooling_2}
\end{figure*}

To achieve this, we consider a scheme similar to  \cite{Shi_cooling_PRA_2023} and introduce a strong AC-Stark shift to decouple the electronic and nuclear degrees of freedom. In~\cite{Shi_cooling_PRA_2023}, a $\pi$-polarized light is used, and for this case, one can only decouple the electronic and nuclear degrees of freedom for a qubit which is encoded in $\ket{0}=\ket{M_I=-9/2}$ and $\ket{1}=\ket{M_I=-7/2}$.
Here we generalize this scheme to any encoding of the nuclear spin states, which can be used for qudit-based quantum computing~\cite{omanakuttan2021quantum} or building error-correcting codes~\cite{Omanakuttan_spincats_PRXQ_2024}.

Consider a resonant coupling between  manifolds  $\mathrm{5s5p ^{1}P_1} \equiv a $ and $\mathrm{5s6s ^1S_0} \equiv b$.   For light linearly polarized along the  $x$-direction
,  the AC-Stark shift interaction Hamiltonian is
\begin{equation}
    \begin{aligned}
        H_{int}=&\frac{\Omega_{ab}}{2\sqrt{2}}\left(\ket{a,M_J=-1}\bra{b,M_J=0}\right.\\
        &\left.-\ket{a,M_J=1}\bra{b,M_J=0}+\mathrm{h.c}\right),
    \end{aligned}
    \label{eq:light_shift_Hamiltonian}
\end{equation}
This leads to an Autler-Townes splitting; the states $M_J=\pm 1$ are light shifted by $\pm \Omega_{ab}/2\sqrt{2}$.  For sufficiently large values of Rabi-frequency $\Omega_{ab}$, the different $M_J$ states are well separated compared to the hyperfine coupling,  and in this regime one can solely access the $M_J=0$ state without transferring population to $M_J=\pm 1$.

The eigenstates of AC-Stark interaction are the Autler-Townes coupled states, denoted 
\begin{equation}
    \{\ket{a,M_J=0,M_I},\ket{\widetilde{M}_J=+,M_I},\ket{\widetilde{M}_J=-,M_I}\}.
\end{equation}
where the dressed states are,
\begin{equation}
\begin{aligned}
    &\ket{\widetilde{M}_J=+}=\frac{1}{\sqrt{2}}\left(\ket{a,M_J=1}-\ket{b,M_J=0}\right),\\
     &\ket{\widetilde{M}_J=-}=\frac{1}{\sqrt{2}}\left(\ket{a,M_J=-1}-\ket{b,M_J=0}\right).
\end{aligned}    
\end{equation}
In the regime of large $\Omega_{ab}$, one can use perturbation theory to find the energy shift to first and second order due to the hyperfine interaction.   Focusing on the states $\ket{a,M_J=0,M_I}$,
 \begin{equation}
 \begin{aligned}
 \delta E_{M_I}^{(1)} &=Q'\left[3\left(I(I+1)-M_I^2\right)+I(I+1)J(J+1)\right],\\
 \delta E_{M_I}^{(2)}&=-\frac{2\sqrt{2}}{\Omega_{ab}} \sum_{M_{I'}} \lvert \bra{M_{I'},M_J=1} H_{\mathrm{hf}}\ket{M_I,M_J=0}\rvert^2,\\
 &+\frac{2\sqrt{2}}{\Omega_{ab}} \sum_{M_{I'}} \lvert \bra{M_{I'},M_J=-1} H_{\mathrm{hf}}\ket{M_I,M_J=0}\rvert^2,
 \end{aligned}
\end{equation}
where,
\begin{equation}
    Q'=\frac{Q}{2I J(2I-1)(2J-1)}.
\end{equation}
The first-order shift $\delta E^{(1)}$ arises from the quadrupolar hyperfine terms and leads to a level shift.  The second-order perturbation, $\delta E^{(2)}$, depends on the strength of the hyperfine coupling compared to the Autler-Townes splitting and can be made negligible with sufficient laser power.

In \cref{fig:cooling_2}, we show the energy shift and overlap of the exact eigenvectors with the perturbative approximation;  we choose here $\Omega_{ab}/2\pi=1$ GHz, achievable with an experimentally reasonable intensity.
In \cref{fig:cooling_2}(a), we compare the exact energy shift to the one obtained from first-order and second-order perturbation theory for state $\ket{M_J=0, M_I}$, showing good agreement.  In \cref{fig:cooling_2}(b), we show an analysis of the eigenvectors and plotted the fidelity,
\begin{equation}
    \mathcal{F}(M_I)=\lvert \bra{n(\Omega_{ab,M_I})}\ket{M_J=0,M_I}\rvert^2,
\end{equation}
where $\ket{n(\Omega_{ab,M_I})}$ is the exact eigenstates for a Rabi-frequency $\Omega_{ab}$ and detuning $\Delta_{ab}$ for the Hamiltonian given in $H=H_{\mathrm{HF}}+H_{\mathrm{LS}}$.
Thus, for on-resonance driving with  $\Omega_{ab}/2\pi=1$ GHz, the eigenvectors are well approximated by the product states, with a little admixture of $M_J =\pm1$, $\ket{M_J=0, M_I}$.
Hence, the good quantum numbers are $M_J, M_I$ rather than $M_F$ and the polarization degree of freedom of the scattered light does not have any information about the nuclear spin state $\ket{M_I}$.  Optical pumping between different $M_I$ in the ground state is strongly suppressed.

 With the polarization dependence of the scattered light removed one needs also to resolve the frequency dependence on scattered light which occurs when we can resolve $|b,M_I\rangle\rightarrow|g,M_I\rangle$ transitions. 
This arises due to the residual light shift dominated by an approximately quadratic hyperfine shift, $\delta_{M_I}= Q'M_I^2$ in the $^1\mathrm{P}_1, M_J=0$ manifold. 
In~\cite{Gorshkov2009}, a strong magnetic field was used to decouple the different $|b,M_J\rangle$ states~\cite{Reichenbach_cooling}, and a large detuning $\Delta\gg Q,\Omega,\Gamma$ was used to negate the frequency-dependent decoherence. 
We show below how this lightshift can be canceled by dressing the $M_J=0$ manifold appropriately, enabling resonant detection without decoherence.


\subsection{Removing the frequency-dependent decoherence}\label{subsec:frequency_decoherence}
\subsubsection{Transfer of coherence in spontaneous emission}
\label{subsec:decoherence_free_photon_scattering}
\begin{figure}
    \centering    
    \includegraphics[width=0.7\columnwidth]{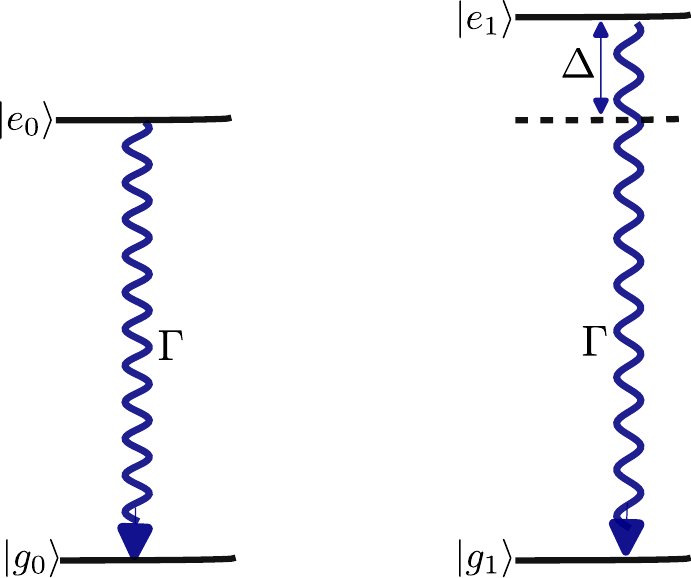}
    \caption{Decoherence during spontaneous decay due to the frequency difference of spontaneously emitted photons. 
    Consider a superposition in the excited states $\ket{e_0}, \ket{e_1}$, which spontaneously decay to $\ket{g_0},\ket{g_1}$ respectively. 
    The two transitions $\ket{e_i}\rightarrow\ket{g_i}$ have the same decay rate $\Gamma$ but differ in transition frequency by $\Delta$. 
    The loss of fidelity due to ``which way information" is transferred to the ground subspace, proportional to $\Gamma^2/(\Gamma^2+\Delta^2)$.
    For a larger $\Delta$, the two emitted photons would be more distinguishable, leading to more decoherence.
}
    \label{fig:two_levels_a}
\end{figure}

The frequency of a spontaneously emitted photon will have information about the spin state of the nucleus when photons emitted on different possible transitions can be spectrally resolved.
To understand this, consider the simplest case of ground and excited states with, two-sublevels. 
In the excited state we include a small energy splitting $\Delta$ and decay rate $\Gamma$; see \cref{fig:two_levels_a}. We can then study the transfer of coherence in the decay of superpositions in the excited states to the ground state. The setting of the problem is given in \cref{fig:two_levels_a} and the Hamiltonian of interest is given as,
\begin{equation}
    H=-\Delta \ketbra{e_1}{e_1}
\end{equation}
in a rotating frame.  The jump operators which take into account the transfer of population and coherence from the excited to the ground state are given as,
\begin{equation}
L=\sqrt{\Gamma}\left(\ket{g}_0\bra{e}_0+\ket{g}_1\bra{e}_1\right),
\label{eq:jump_two_levels_a}
\end{equation}
and the  evolution of the system is given by the Lindblad master equation,
\begin{equation}
    \frac{d\rho}{dt}=-i\comm{H}{\rho}-\frac{\Gamma}{2}\left(L^{\dagger}L \rho+ \rho L^{\dagger}L \right)+ \Gamma L\rho L^{\dagger}.
    \label{eq:Lindblad_master_equation}
\end{equation}
To identify the loss of coherence, consider, 
\begin{equation}
        \rho_{0,1}^{(g)}\equiv\bra{g_0} \rho \ket{g_1},\; \; \rho_{0,1}^{(e)}\equiv\bra{e_0} \rho \ket{e_1}.
\end{equation}
Thus,
\begin{equation}
\frac{d\rho_{0,1}^{(e) }}{dt}=-(i\Delta+\Gamma) \rho_{0,1}^{(e)},
\end{equation}
which in turn gives,
\begin{equation}
\rho_{0,1}^{(e)}(t)=\rho_{0,1}^{(e)}(0)e^{-(i\Delta+\Gamma)t}.
\end{equation}
Similarly, 
\begin{equation}
\begin{aligned}
   \frac{d\rho_{0,1}^{(g) }}{dt}&=\Gamma\rho_{0,1}^{(e)}, \\
   \frac{d\rho_{0,1}^{(g) }}{dt}&=\Gamma\rho_{0,1}^{(e)}(0)e^{-(i\Delta+\Gamma)t}. 
\end{aligned}
\end{equation}
Solving the above differential equation,
\begin{equation}
\rho_{0,1}^{(g)}(t)=\frac{\Gamma}{\Gamma-i\Delta}\left(1-e^{-(i\Delta-\Gamma)t}\right)\rho_{0,1}^{(e)}(0),
\end{equation}
which in limit of $t\gg 1/\Gamma$ is,
\begin{equation}
    \rho_{0,1}^{(g)}(t) \approx \frac{\Gamma}{\Gamma-i\Delta} \rho_{0,1}^{(e)}(0).
\end{equation}
To understand the effect consider an initial state,
\begin{equation}
    \ket{\psi}_0=\sqrt{\frac{1}{2}}\left(\ket{e_0}+\ket{e_1}\right),
\end{equation}
and consider the evolution for a total time $T=10\Gamma$ and for an ideal transfer of population one expects to get the state,
\begin{equation}
    \ket{\psi}_f=\sqrt{\frac{1}{2}}\left(\ket{g_0}+\ket{g_1}\right),
    \label{eq:initial_state_app_1}
\end{equation}
and we can calculate the fidelity under the evolution by the \cref{eq:Lindblad_master_equation} given as,
\begin{equation}
    \mathcal{F}=\bra{\psi}_f\rho(t)\ket{\psi}_f.
    \label{eq:fidelity_two_levels_a}
\end{equation}
\begin{figure}[!ht]
    \centering   
    \includegraphics[width=1\columnwidth]{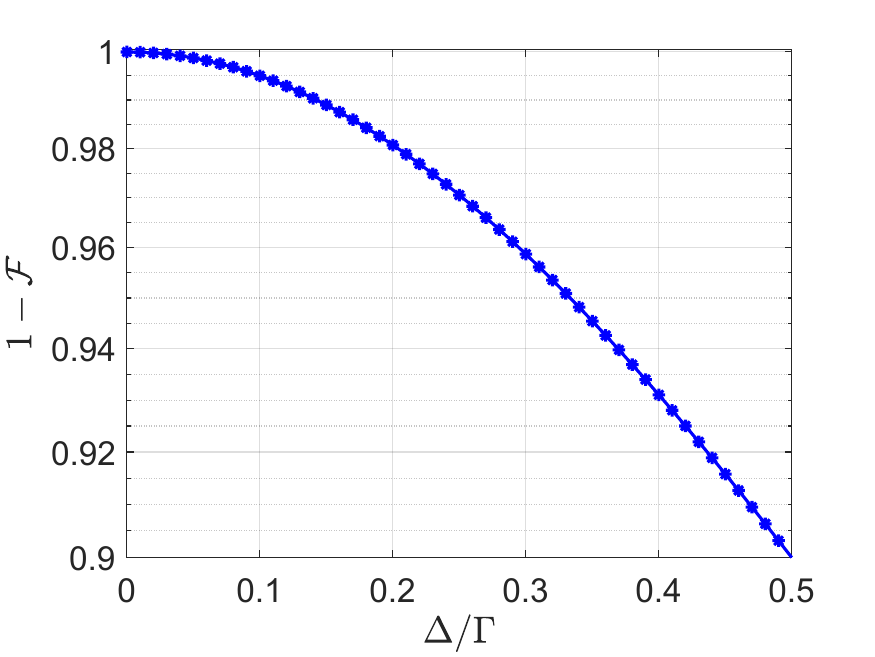}
    \caption{Infidelity of the final state due to the loss of transfer of coherence in spontaneous emission as a function of $\Delta/\Gamma$ for two two-level systems and  \cref{eq:fidelity_two_levels_a}. 
    As $\Delta/\Gamma\to 0$, the infidelity goes to $0$, and hence the coherence is preserved in the decay.
}
    \label{fig:two_levels_fidelity_a}
\end{figure}

In \cref{fig:two_levels_fidelity_a}, we show the infidelity as a function of $\Delta/\Gamma$ and it is evident that we have good fidelity as  $\Delta/\Gamma \to 0$.
Thus to overcome the frequency dependence for decoherence-free scattering near resonance, one needs to ensure that the energy difference between all magnetic sublevels is much less than the atomic linewidth.

\subsubsection{Removing which-way information in the photon's frequency}
\begin{figure*} 
    \centering
    \includegraphics[width=2\columnwidth]{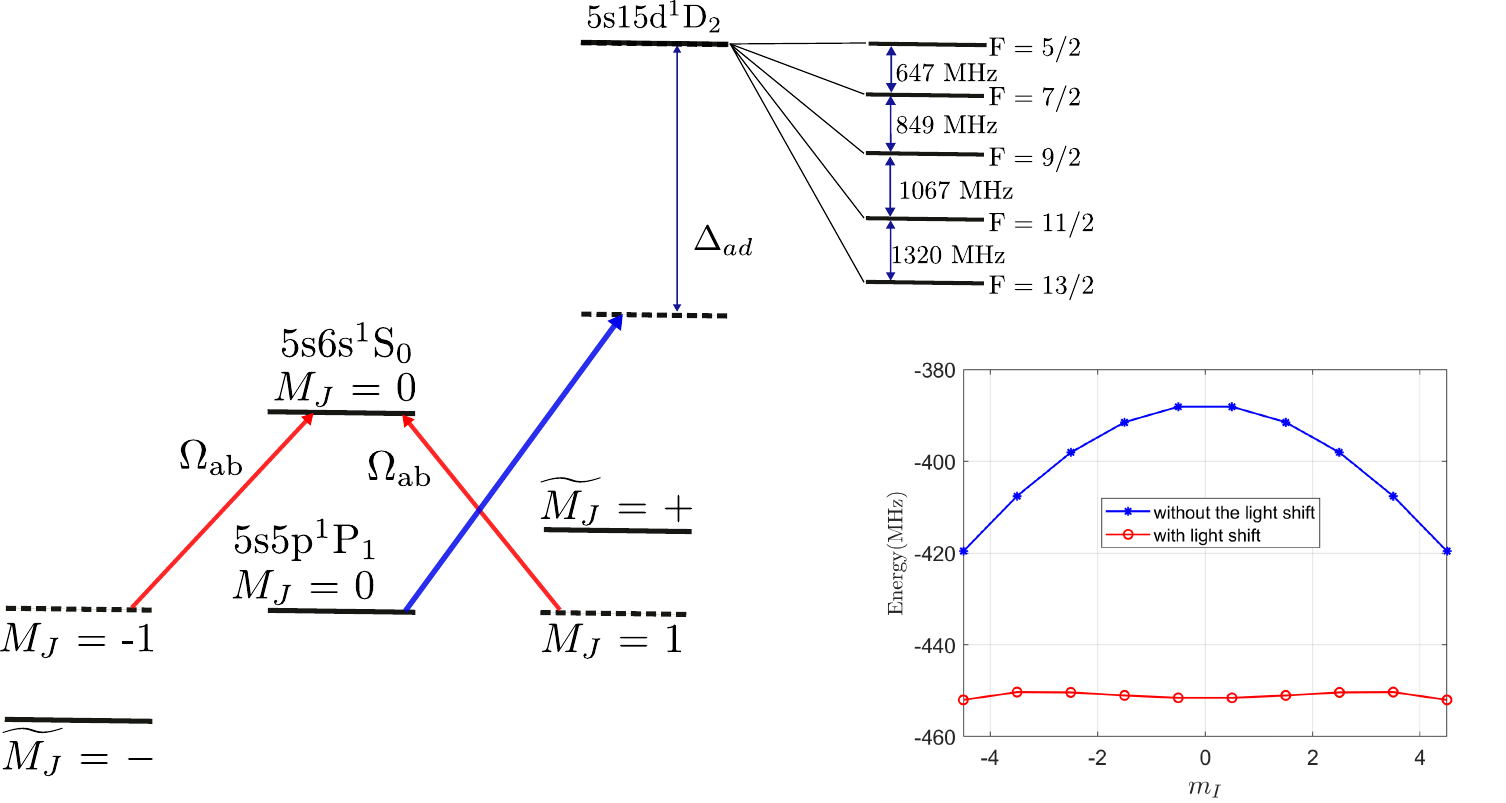}
    \caption{ Canceling the quadrupolar hyperfine shift. 
    A strong $\sigma$-polarized laser coupling the $5s5p ^1\mathrm{P}_1,M_J=\pm1$ to the 5s6s $^1\mathrm{S}_0$ states induces a strong light shift that dominates over the hyperfine interaction energy. The product basis becomes the good quantum numbers, and the
    new eigenstates in the $M_J=0$ manifold become pure nuclear states (see \cref{fig:S_dressing}). 
    These nuclear spin states have a residual quadrupolar hyperfine shift as shown with blue stars in the inset. We cancel this level shift by dressing the $^1\mathrm{P}_1, M_J=0$ manifold with 5s15d$^1D_2$ manifold.
    The detuning $\Delta_{ad}$ is chosen such that the resulting tensor light shift optimally cancels the quadrupolar hyperfine shift, as shown in red dots in the inset.  }
    \label{fig:cooling_setup}
\end{figure*}

Due to the large quadrupolar level shifts of the nuclear spin states for the $\mathrm{^1P_1}$ manifold, spontaneous emission will lead to loss of coherence between sublevels that are well separated compared to the excited-state linewidth.  
To remove this residual energy shift, we consider the application a tensor light shift by off-resonantly coupling the  $\mathrm{5s5p^1P_1}, M_J=0$ state to the  $\mathrm{5s15d^1D_2}\equiv d$ manifold.  
For $\pi$-polarized light with electric field amplitude $\mathcal{E}_{\bm{d}}$, the light state for state $\ket{a,M_J=0,M_I}$ is 

\begin{equation}
    V_{\mathrm{LS}}^{(ad)}= \sum_{F',M_{F'}}\frac{\mathcal{E}^2_{\bm{d}}}{4\Delta_{a,dF'}}\left\lvert\langle{d,F',M_{F'}}\lvert \bm{d}_z\ket{a,M_J=0,M_I}\right\rvert^2,
    \label{eq:light_shift_Hamiltonian_d}
\end{equation}
where $\Delta_{a,dF'}$ is the detuning between the $\mathrm{5s5p^1P_1, M_J=0}$ state and the hyperfine manifold $F'$ of $\mathrm{5s15d^1D_2}$ . 
Working near resonance for $F'=13/2$, using the fact that the  dipole allowed interaction only allows  $F=F'\pm 1$, and for $\pi$-polarization $M_{F'}=M_I$, the only matrix element we need to consider is,
\begin{widetext}
    \begin{equation}
    \begin{aligned}
        &\bra{d,F'=13/2,M_I}\bm{d}_z\ket{a,1,0;9/2,M_I}        =\bra{d,F'=13/2,M_I}\bm{d}_z\ket{a,F=11/2,M_I}  \braket{F=11/2,M_I}{1,0;9/2,M_I},\\
        &= \rme{d,F'}{\bm{d}_z}{a,F}\braket{F'=13/2,M_I}{1,0;F=11/2,M_I}\braket{F=11/2,M_I}{1,0;9/2,M_I},\\
        &=\mathcal{O}^{J',F}_{J,F}\rme{d,F'}{\bm{d}_z}{a,F}\braket{F'=13/2,M_I}{1,0;F=11/2,M_I}\braket{F=11/2,M_I}{1,0;9/2,M_I},
    \end{aligned}
    \label{eq:knowing_the_quartic_dependence_2}
\end{equation}
\end{widetext}
having used the Wigner-Ekart theorem with $\rme{d,F'}{\bm{d}_z}{a,F}$ the reduced dipole matrix element.  $\mathcal{O}^{J',F'}_{J,F}$ is the relative oscillator strength defined as~\cite{Deutsch2000},

 \begin{equation}
   \mathcal{O}^{J',F'}_{J,F}=(-1)^{F'+1+F+I}\sqrt{(2J'+1)(2F+1)}\begin{Bmatrix}
       F' & I & J'\\
       J & 1 &F
   \end{Bmatrix}, 
\end{equation}

\noindent and for the case of $J'=2,J=1,F'=13/2,F=11/2$, we get $\mathcal{O}^{J',F'}_{J,F}=1$.
Using the following  property of Clebsch-Gordan coefficients \cite{ME_Rose_angular_momentum},
\begin{equation}
    \braket{j+1,m}{1,0;j,m}=\sqrt{\frac{(j+1)^2-m^2}{(2j+1)(j+1)}},
\end{equation}
 gives,
\begin{equation}
    \begin{aligned}
        &\braket{F'=13/2,M_I}{1,0;F=11/2,M_I}=\frac{1}{2}\sqrt{\frac{169-4M_I^2}{78}}\\
        &\braket{F=11/2,M_I}{1,0;I=9/2,M_I}=\frac{1}{2}\sqrt{\frac{121-4M_I^2}{55}}.
    \end{aligned}
\end{equation}
Thus, the contribution of the tensor light-shift interaction when detuned from the resonance $F'=13/2$  is,
\begin{equation}
\begin{aligned}
   V_{\mathrm{LS}}^{ad}= &\lvert \bra{d,F',M_{F'}} \bm{d}_z \ket{a,J,M_J;I,M_I} \rvert^2\\
    &=V_0^{ad}\left(0.298-0.0169M_I^2+0.000233M_I^4\right).
    \end{aligned}
\end{equation}
where $V_0^{ad}=\frac{\Omega_{ad}^2}{4\Delta_{ad}}\lvert\langle 2,0  \rvert  1,0;1,0\rangle \rvert^2$, for Rabi frequency
$\Omega_{ad} = \langle d,F' ||\bm{d}|| a,F \rangle \mathcal{E}_d$. 
The quartic behavior is not familiar for the light which is traditionally at most quadratic and $\Delta_{ad}$ is the detuning as shown in \cref{fig:cooling_setup}.  
This arises here from the coupling to the electron through $J=2$ (which is quadrupolar rather than dipolar). 
For further details, see \cref{sec:tensor_light_shift}.   
The dominant quadratic term can be used to cancel the energy light arising from the hyperfine perturbation in the state $a$, which also has a quadratic term from the perturbation theory analysis.

 Figure~(\ref{fig:light_shift_spectrum}a) shows the shifts of the magnetic sublevels in $M_J=0$ manifold of the $\mathrm{^1P_1}$ state in the presence and absence of the off-resonant light shift interaction.  The additional light shift effectively cancels the residual quadrupolar hyperfine shift which highly suppresses the which way information about the nuclear spin state in spontaneous emission.  
  
\section{Decoherence-Free Resolved-Sideband Cooling}
\label{sec:qnd_cooling}

\begin{figure}
    \centering
    \includegraphics[width=0.7\columnwidth]{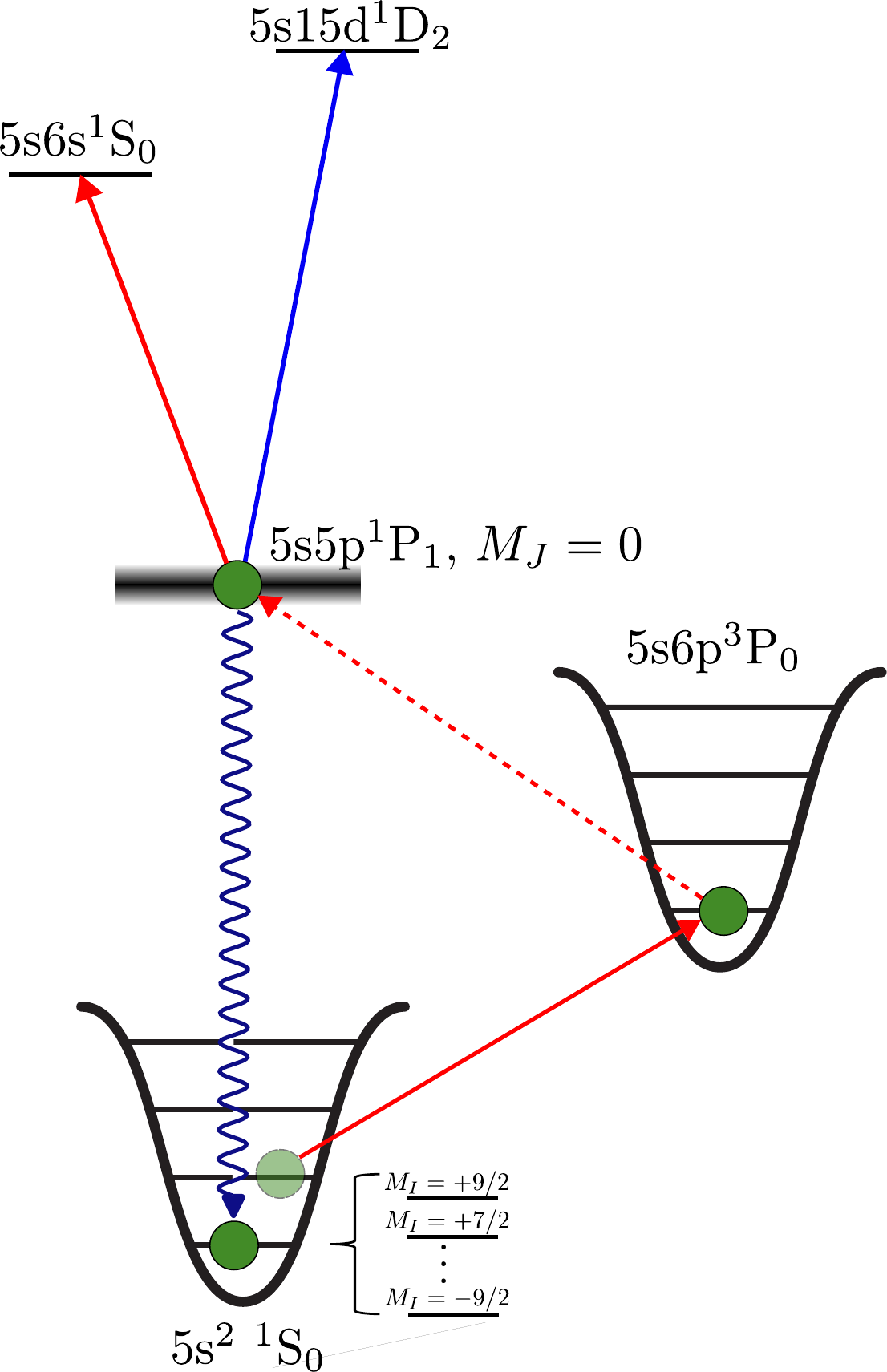}
    \caption{ Nuclear spin coherence-preserving sideband cooling.
   Population is transferred from a vibrational level $n$ in the ground $^1\mathrm{S}_0$ state to  vibrational level $n-1$ in the clock $^3\mathrm{P}_0$ state via excitation on the red sideband. 
    The population is then quenched by pumping into the $^1\mathrm{P}_1$ state where it rapidly decays into the ground $^1\mathrm{S}_0$ with vibration quantum number $n-1$, when in the Lamb-Dicke regime. Due to the dressing by the 5s6s$^1\mathrm{S}_0$ and the 5s15d$^1\mathrm{D}_2$ manifolds, the nuclear spin coherence is preserved during this decay.}
    \label{fig:cooling}
\end{figure}

The QND leakage detection through resonance fluorescent will have the adverse effect of recoil heating of the atomic motion.  While this can be somewhat mitigated through tight trapping and the Lamb-Dicke effect~\cite{Lamb_Dicke, Enders_Lamb_Dicke, Kaufman_Lamb_Dicke}, the accumulated heating will lead to increased errors in quantum information processing.  Such heating will inevitably arise as well in protocols involving atomic transport and in entangling gates where atoms are released from traps and recaptured.  It would thus be advantageous to combine the protocol that performs QND leakage-to-erasure conversion with one that also cools the atom motion.  We can achieve this with the tools discussed above, and integrate them into a protocol that implements resolved sideband cooling without decohering the nuclear spin, as originally proposed in~\cite{Reichenbach_cooling}.  
In that work, a large magnetic field was considered in order to decouple the nuclear spin from the electronic degrees of freedom.
Here, we consider the efficacy of sideband cooling by employing the AC-Stark shift as discussed in \cref{sec:qnd_leakage_detection}, extending the work of ~\cite{Shi_cooling_PRA_2023} to the full ten-dimension qudit manifold, for the $I=9/2$ nuclear spin in $^{87}$Sr.

The resolved-sideband cooling for $^{87}$Sr follows three steps, as in  previous works \cite{Reichenbach_cooling,Shi_cooling_PRA_2023},  and shown in \cref{fig:basic_outline}. 
In the first step using a $\pi$-pulse we coherently excite 
\begin{equation}
    \begin{aligned}
        &\ket{\mathrm{5s^2\text{ } ^1S_0}, M_I}\otimes \ket{n} \\
        \to& \ket{\mathrm{5s5p\text{} ^3P_0},M_F=M_I}\otimes \ket{n-1},
    \end{aligned}
\end{equation}
on the first red sideband, where $n$ is the vibrational quantum number. 
In the next step, using a two-photon transition, we coherently transfer 
\begin{equation}
    \begin{aligned}
        &\ket{\mathrm{5s5p\text{} ^3P_0},M_F=M_I} \otimes \ket{n-1} \\ \to &\ket{\mathrm{5s5p\text{} ^1P_1},M_J=0,M_F=M_I},\otimes \ket{n-1}.
    \end{aligned}
\end{equation}
in order to rapidly repump the population.  Assuming a sufficiently tight trap in the Lamb-Dicke regime, in the last step, the short lifetime  of the $\mathrm{5s5p\text{ } ^1P_1}$ leads to the decay
\begin{equation}
\begin{aligned}
    &\ket{\mathrm{5s5p\text{} ^1P_1},M_J=0,M_F=M_I} \otimes \ket{n-1}\\
    \to &\ket{\mathrm{5s^2\text{ } ^1S_0}, M_I}\otimes \ket{n-1},
\end{aligned}    
\end{equation}
thus returning the population to the ground state but with a decreased vibrational quantum number.  The schematic is given in \cref{fig:cooling}.

\begin{figure}
    \centering
    \includegraphics[width=1\columnwidth]{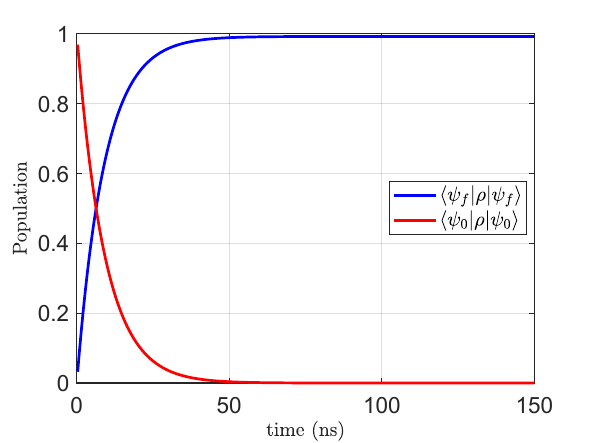}
    \caption{
    An initial state of equal nuclear spin superposition (\cref{eq:initial_state_cooling})  is considered in the $^1\mathrm{P}_1$ manifold for the scheme considered in \cref{fig:cooling_setup}.
        We simulate the dynamics under the Lindblad master equation. 
    The fidelity of the final state with respect to the equal superposition state in the ground $^1\mathrm{S}_0$ manifold (\cref{eq:final_state_cooling}). 
 }
    \label{fig:light_shift_spectrum}
\end{figure}


The key to decoherence-free cooling is that the nuclear spin is a spectator throughout the cycle.   
The  $\mathrm{^3P_0}$ clock state has negligible hyperfine coupling, and the nuclear spin state can be transferred from the ground to the clock state manifold in step-one while preserving all nuclear-spin coherence.
In step-two we employ the addition of resonant laser field on the $\mathrm{5s5p ^1P_1} \rightarrow \mathrm{5s6s ^1S_0}$ and off-resonant $\mathrm{5s5p ^1P_1} \rightarrow \mathrm{5s15d ^1D_2}$ transitions to both decouple the electron and nuclear spin and remove the large quadrupolar nuclear level shift, as discussed in \cref{sec:qnd_leakage_detection}.  
To quantify how the nuclear spin coherence is conserved in the cooling cycle, we consider an initial state,
\begin{equation}
\ket{\psi}_{0}
     =\frac{1}{\sqrt{10}} \sum_{M_I=-\frac{9}{2}}^{\frac{9}{2}} \ket{a,M_J=0,M_I}.
     \label{eq:initial_state_cooling}
 \end{equation}
Evolving under the master equation with the Lindblad jump operator
    \begin{equation}
    L=\sqrt{\Gamma} \sum_{i-\frac{9}{2}}^{\frac{9}{2}}\ket{g,M_J=0,M_I=i}\bra{a,M_J=0,M_I=i},
\end{equation}
we calculate the the population in the desired final state
\begin{equation}
    \ket{\psi}_{f}
     =\frac{1}{\sqrt{10}} \sum_{M_I=-\frac{9}{2}}^{\frac{9}{2}} \ket{g,M_J=0,M_I}.
     \label{eq:final_state_cooling}
\end{equation}
Figure \ref{fig:light_shift_spectrum} shows the time evolution for $\Omega_{ab}/2\pi=1$ GHz, $\Omega_{ad}/2\pi=106$ MHz, and $\Delta_{ad}/2\pi=4350$ MHz.  
We have optimized the $\Omega_{ad}$  and $\Delta_{ad}$ such that the frequency dependence of the spontaneously emitted photons is lowest for these values. From the figure one can see that the excited state decays to the ground state and for the choice of parameter considered the coherence is completely preserved during the decay of the information to the ground state, thus giving us a high-fidelity QND cooling scheme.  The final temperature of the atoms will depend on the Lamb-Dicke parameter and control of excitations of the red-sideband and repumping.

\section{Summary and Conclusion}
\label{sec:conclusions_and_future_work}

In this work, we propose methods to tackle leakage, atom loss, and atom heating in alkaline-earth (like) neutral atom quantum information processors. These errors are a major hurdle for high-fidelity operations with long circuit depths and active error correction rounds.  Leakage will arise when the population is left behind in other levels beyond the computational subspace. Atoms may be lost from the trap due to background collisions, anti-trapping of atoms left in Rydberg states, or during trap turn-off and recapture for entangling operations. Similarly, atoms can heat up during entangling operations, or atom shuttling. These effects will build up over the course of many operations. While these problems could be mitigated using ancilla atoms and circuit-based methods~\cite{chow2024_erasure}, the methods discussed here do not require any extra qubit overhead and can work for non-reconfigurable arrays.

 Our leakage/loss QND detection and cooling is made possible by the rich atomic structure of alkaline-earth (like) atoms. The existence of electron $J=0$ states allow one to store quantum information in the isolated nuclear spin.  Through laser coupling of the electrons to states with $J\neq 0$, one can manipulate qubits (or more general qudits) encoded in the nuclear spin through the hyperfine interaction.  On the flip side, if one can suppress the hyperfine interaction, one can manipulate the atom without disturbing quantum information stored in the nucleus, which is solely a spectator in the laser-electron interaction.  We use this for QND leakage detection of qudits stored in the ground state of $^{87}$Sr, which can also be accompanied by simultaneous cooling of the atoms


We first present methods to detect the population in the ground $^1\mathrm{S}_0$ manifold without nuclear spin decoherence through fluorescence on the
$^1\mathrm{S}_0\leftrightarrow ^1\mathrm{P}_1$ transition,
with a laser field far-off resonance compared to the small hyperfine splitting in the excited state.
For such detuning  $\Delta_{ga}$, the ratio of scattering rate to optical pumping falls off as $1/\Delta_{ga}^2$, and the resulting elastic Rayleigh scattering enables us to detect the atom without optical pumping.
The off-resonant field induces a residual coherent tensor light shift, but this can be reversed by the opposite shift induced by coupling to the long-lived $^3\mathrm{P}_1$ state. 
In a second method, we suppress the hyperfine coupling in the $^1\mathrm{P}_1$ manifold by decoupling the nuclear spin from the electron using a strong AC Stark shift. 
We isolate the $^1\mathrm{P}_1, M_J=0$ state be resonantly
dressing the $^1\mathrm{P}_1, M_J=\pm 1$ states with the 
excited $6s^2\text{ } ^1S_0$ state.
The residual quadrupolar hyperfine shift on $M_J=0$ is canceled by dressing with the 5s15d$^1\mathrm{D}_2$ manifold. 
With this one can resonantly scatter light from the $^1\mathrm{P}_1,M_J=0$ manifold with very little nuclear spin decoherence.

 Finally, we show how one can combine QND leakage detection from the ground state and cooling an atom without nuclear spin decoherence. 
 This will be an important tool, as atoms heat up over the course of many operations, increasing the probability of atom loss and decreasing gate fidelities. 
 The ground $^1\mathrm{S}_0$ and the clock $^3\mathrm{P}_0$ manifolds have zero electron angular momentum $J=0$ and hence no hyperfine coupling.
 In combination with the hyperfine-decoupled $^1\mathrm{P}_1, M_J=0$ manifold, they form a subspace where the electrons are effectively decoupled from the nucleus. 
 We can now cool the atoms by red-sideband driving them to the clock manifold and then quenching by transferring them to the $^1\mathrm{P}_1$ manifold.
 This is made possible due to the long lifetime of clock states and the motional mode selectivity of the clock transition.  Our protocol is a generalization that combines elements of previously studied cooling schemes~\cite{Reichenbach_cooling, Shi_cooling_PRA_2023}.

In the main text, we considered encoding information in the nuclear spin of $^{87}$Sr ($I=9/2$); we studied $^{171}$Yb ($I=1/2$) in the Appendix.
The nuclear spin-1/2 of $^{171}$Yb makes it a natural qubit, and avoids phenomena such as tensor light shifts and quadrupolar hyperfine shifts, but comes at the cost of large magnetic dipole hyperfine couplings.  Our methods are also applicable to other fermionic alkaline-earth (like) atoms for e.g., $^{173}$Yb (I=5/2). These methods also have synergy with the spin-cat encoding~\cite{Omanakuttan_spincats_PRXQ_2024}. Together, these techniques could significantly advance the prospects of fault-tolerant computations with alkaline-earth (like) neutral atoms.


\vspace{0.2cm}

\begin{acknowledgements}
The authors acknowledge fruitful discussions with Anupam Mitra and Milad Marvian during various stages of this work. 
We also thank Alexey Gorshkov for the helpful discussions.
This work was supported by the Laboratory Directed Research and Development program of Los Alamos National Laboratory under project numbers 20240295ER and 20210116DR,
and the NSF Quantum Leap Challenge Institutes program, Award No. 2016244.

This paper was prepared for informational purposes with contributions from the Global Technology Applied Research center of JPMorgan Chase \& Co.
This paper is not a product of the Research Department of JPMorgan Chase \& Co. or its affiliates. Neither JPMorgan Chase \& Co. nor any of its affiliates makes any explicit or implied representation or warranty, and none of them accept any liability in connection with this position paper, including, without limitation, with respect to the completeness, accuracy, or reliability of the information contained herein and the potential legal, compliance, tax, or accounting effects thereof. This document is not intended as investment research or investment advice, or as a recommendation, offer, or solicitation for the purchase or sale of any security, financial instrument, or financial product or service, or to be used in any way for evaluating the merits of participating in any transaction.
\end{acknowledgements}

\appendix

\clearpage
\section{QND leakage detection in $^{171}$Yb}
\label{sec:qnd_leakage_detection_yb}
\begin{figure*}[!ht]
        \centering   
    \subfloat[]{\includegraphics[width =0.84\columnwidth]{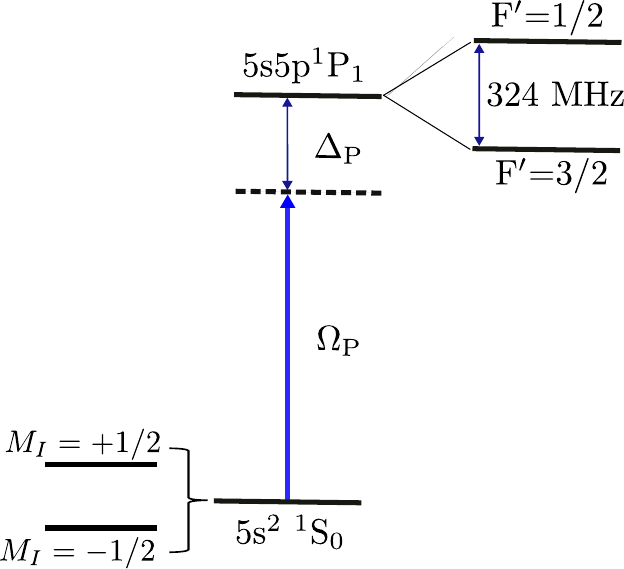} \label{fig:qnd_leakage_yb_a}}\hspace*{1 em}
    \subfloat[]{\includegraphics[width =1.16\columnwidth]{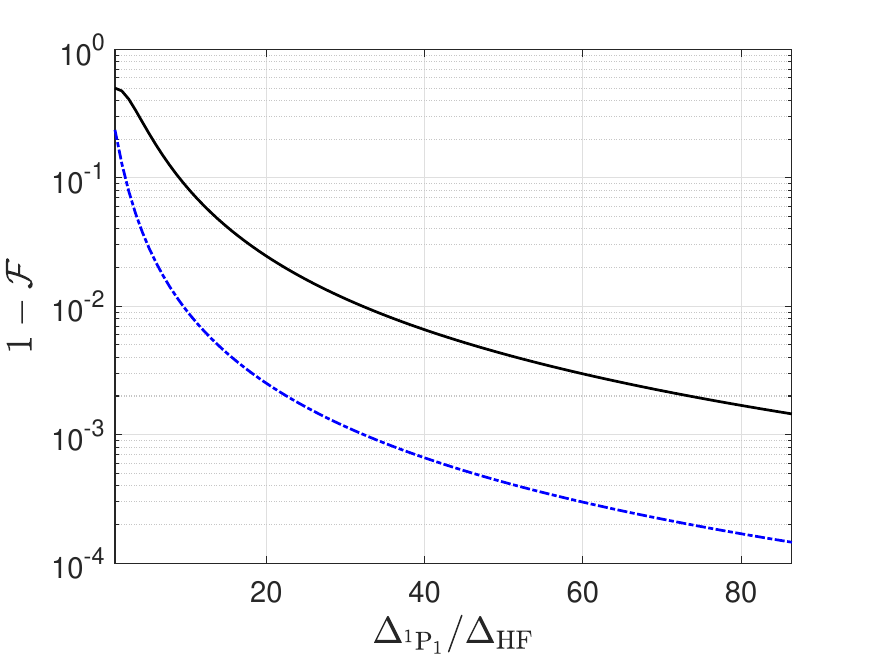}  \label{fig:qnd_leakage_yb_b}}
       \caption{The figure illustrates the setup for detecting the loss of information in the state encoded in the ground state of $^{171}$Y b.
    In (a) shows the level diagram, we utilize far-detuned light from the singlet P state ($\mathrm{6s6p^1P_1}$). 
    For the case of the Yb, there is no tensor light-shift and thus we only need a single laser for detecting the loss of atoms.
    Since the hyperfine splitting is large compared to the case of Sr, we need to further go off-resonance for a perfect QND leakage detection scheme.
    (b) show the simulation of infidelity as a function of detuning from the singlet state for the state given in \cref{eq:initial_state_leakage_yb}.
    The solid(dashed) line is the case when we do the simulation for a time required for scattering $100$ ($10$) photons respectively. 
Lower infidelity indicates a more effective  QND scheme for leakage detection.
Moving further away from resonance enhances the scheme's effectiveness, approaching an ideal scenario for QND leakage detection.
Moving further away from resonance enhances the scheme's effectiveness, approaching an ideal scenario for QND leakage detection.
However, compared to the case of $^{87}$Sr, we need to go further off-resonance for a near-ideal QND leakage detection.
}
    \label{fig:basic_outline_leakage_yb}
\end{figure*}

In this section, we outline the extension of the QND leakage detection scheme for $^{171}$Yb, and the level diagram is given in \cref{fig:qnd_leakage_yb_a}.
To understand the working of the leakage detection scheme, one can study the Lindblad Master equation given as,
\begin{equation}
\frac{d\rho}{dt}=-i\comm{H_{\mathrm{LS}}}{\rho}+\sum_q W_q\rho W_q^{\dagger}-\frac{1}{2}\{W_q^{\dagger}W_q,\rho\}.
\end{equation}
The light shift Hamiltonian in the far-off resonance for a $\pi$ polarized light is given as,
\begin{equation}
H_{\mathrm{LS}} \approx \frac{\Omega_\mathrm{P}^2}{4 \Delta_\mathrm{P}} \left(\mathds{1}-\frac{1}{\Delta_\mathrm{P}}\beta^{(0)} \mathds{1}\right).
\end{equation}
The tensor light shift goes to zero as the nuclear spin in the ground state is $1/2$ and there can only be scalar and vector terms and for $\pi$ polarized light, the vector term also goes to zero.
Thus the jump operators are given as,
\begin{equation}
\begin{aligned}
W_0&\approx \frac{\Omega_{\mathrm{P}}}{2\Delta} \mathds{1}+\frac{\Omega_{\mathrm{P}}}{2\Delta_{\mathrm{P}}^2}\gamma^{(0)} \mathds{1},\\
W_{+}&\approx \frac{\Omega_{\mathrm{P}}}{2\Delta_{\mathrm{P}}^2} \left(i\gamma^{(1)}F_{-}\right),\\
W_{-}&\approx \frac{\Omega_{\mathrm{P}}}{2\Delta_{\mathrm{P}}^2} \left(i\gamma^{(1)}F_{+}\right),
\end{aligned}
\end{equation}
Thus upto third order in $1/\Delta_{\mathrm{P}}$ we get $d\rho/dt \to 0$.
To understand the performance of the scheme in \cref{fig:qnd_leakage_yb_a}, we study the evolution of the  state 
\begin{equation}
\ket{\psi}=\frac{1}{\sqrt{2}}\sum_{i=-\frac{1}{2}}^{\frac{1}{2}} \ket{M_I=i},
\label{eq:initial_state_leakage_yb}.
\end{equation}
The fidelity of the final state is given as,
\begin{equation}
\mathcal{F}=\bra{\psi}\rho\ket{\psi}.
\label{eq:fidelity_leakage_yb}
\end{equation}
In \cref{fig:qnd_leakage_yb_b}, we investigated the fidelity of the state described in \cref{eq:initial_state_leakage_yb} after the time required for detecting $100$ and $10$ photons. Numerical analysis suggests that increasing $\Delta_{\mathrm{P}}$ improves fidelity, and for sufficiently large detunings, the ideal fidelity can be recovered, establishing a QND leakage detection scheme.
Compared to the $^{87}$Sr for the $^{171}$Yb, we need to go further off-resonance for a near-ideal QND leakage detection.

\section{QND Cooling in $^{171}$Yb}
\label{sec:qnd_cooling_yb}

\begin{figure*}[!ht]
         \centering   
    \subfloat[]{\includegraphics[width =0.85\columnwidth]{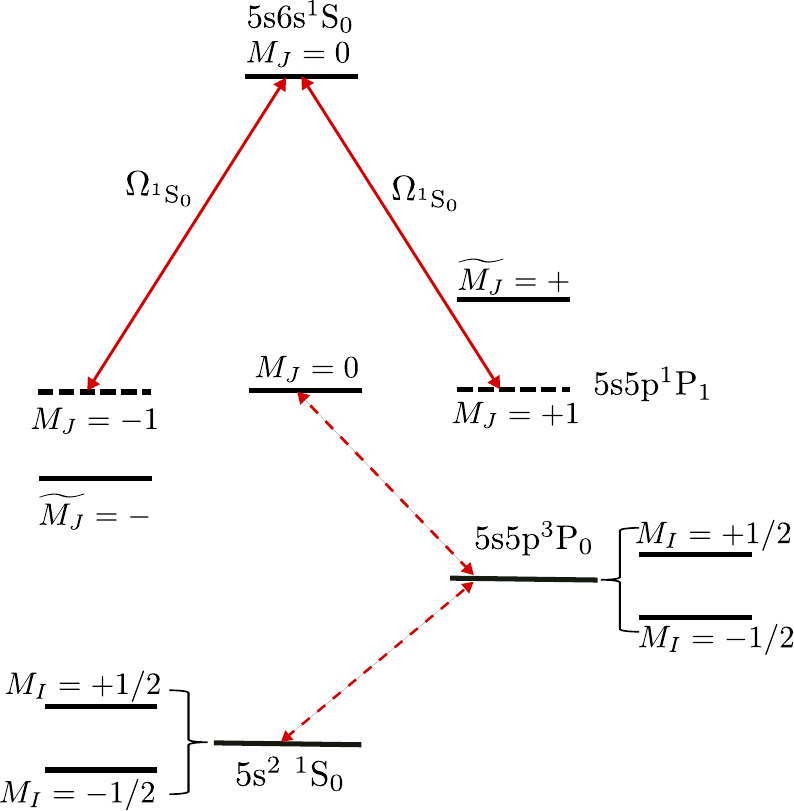}\label{fig:cooling_yb_a}} \hspace*{1 em}
    \subfloat[]{\includegraphics[width =1.15\columnwidth]{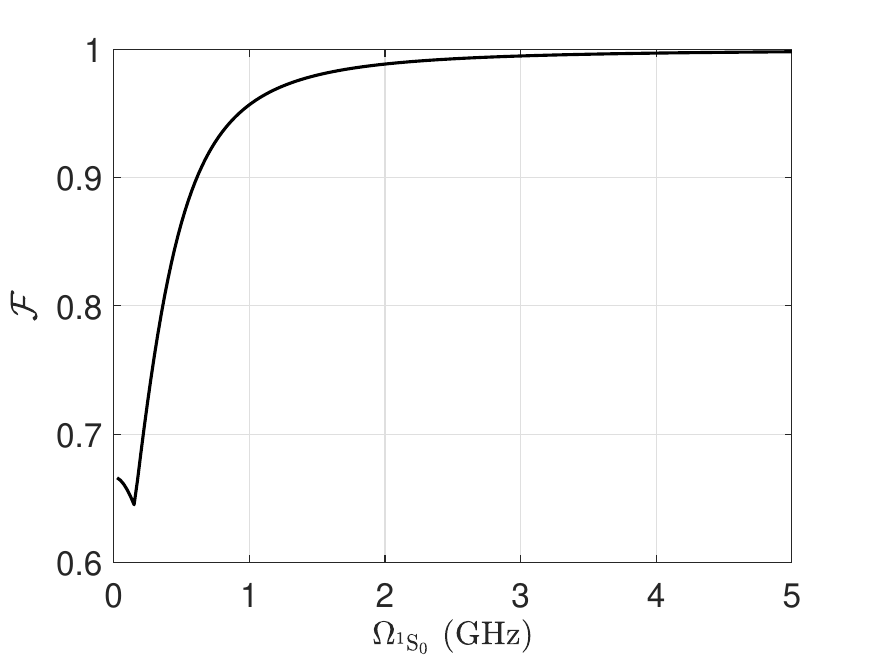}  \label{fig:cooling_yb_b}}
    \caption{The figure illustrates the setup for the QND cooling scheme for a state encoded in the ground state of $^{171}$Yb.
   (a) shows the level diagram, the key ingredient that allows us to overcome the hyperfine splitting interaction in the excited state $\mathrm{6s6p ^1P_1}$, we use AC stark shift to isolate the $M_J=0$ state in this state, to achieve this we couple the $\mathrm{6s6p ^1P_1}$ to the excited singlet state $\mathrm{6s7s ^1S_0}$ using light polarized along the $x$-axis.
   In (b) we calculate $\mathcal{F}$ given in \cref{eq:qnd_cooling_yb_fidelity}, as a function of the Rabi-frequency, and one obtains an ideal QND cooling scheme for sufficiently large Rabi-frequency.
Also, the Rabi-frequency one needs for $^{171}$ Yb is larger than the case of $^{87}$Sr and this is due to the larger Hyperfine interaction for $^{171}$Yb compared to $^{87}$Sr.
}
    \label{fig:cooling_yb}
\end{figure*}
In this section, we outline the extension of the QND cooling scheme for $^{171}$Yb atoms and the schematic is given in \cref{fig:cooling_yb_a}.
Again the key to separate the nuclear and electronic degrees of freedom in the singlet $P$ state. 
As the nuclear spin of $^{171}$Yb is $1/2$, the scheme is much simpler compared to the case of $^{87}$Sr where the nuclear spin is $9/2$.
To achieve a QND cooling scheme, we consider a resonance coupling between $a=\mathrm{6s6p^{1}P_1}$ and $b=\mathrm{6s7s^1S_0}$. 
For a  light polarized along $x$-direction, and the interaction Hamiltonian is
\begin{equation}
    \begin{aligned}
        H_{\mathrm{LS}}=\frac{\Omega_{\mathrm{^1S_0}}}{2\sqrt{2}}&\left(\ket{a,M_J=-1}\bra{b,M_J=0} \right.\\
        &\left. -\ket{a,M_J=1}\bra{b,M_J=0}+\mathrm{h.c}\right),
    \end{aligned}
    \label{eq:light_shift_Hamiltonian_app}
\end{equation}
This leads to an Autler-Townes splitting,  the states $M_J=\pm 1$ are light shifted by $\pm \Omega_{\mathrm{^1S_0}}/2\sqrt{2}$.
For sufficiently large values of Rabi-frequency $\Omega_{\mathrm{^1S_0}}$ the different $M_J$ states are separated and in this regime, one can solely access the $M_J=0$ state without transferring population to $M_J=\pm 1$.
In this regime, the eigenvectors are well approximated by the product state, with a little admixture of $M_J =\pm1$, $\ket{M_J=0,M_I}$ and we have separated the nuclear and electronic degrees of freedom.
Hence, the good quantum numbers are $M_J,M_I$ rather than $M_F$ and the polarization degree of freedom of the scattered light does not have any information about the nuclear spin state $\ket{M_I}$.
However unlike the case of $^{87}$Sr, there is no energy shift for the states $M_J=0,M_I$, and thus there is no frequency dependence of the spontaneously emitted photon giving a  “which way information” about the nuclear spin state.
 Thus one can achieve a perfect QND cooling scheme if we have a sufficiently large Rabi frequency between $a$ and $b$.
 To understand the value of Rabi frequency one needs for a good QND cooling scheme one can study the fidelity of the states to the $M_J=0,M_I$ states and can calculate,
 \begin{equation}
 \begin{aligned}
\mathcal{F}=\frac{1}{2} &\left(\lvert \braket{n(\Omega_{{\mathrm{^1S_0}},M_I=\frac{1}{2}})}{a,M_J=0,M_I=\frac{1}{2}}\rvert^2\right. \\
&\left.+\lvert\braket{n(\Omega_{{\mathrm{^1S_0}},M_I=-\frac{1}{2}})}{a,M_J=0,M_I=-\frac{1}{2}} \rvert^2\right)
\label{eq:qnd_cooling_yb_fidelity}
\end{aligned}
 \end{equation}
where $\ket{n(\Omega_{{\mathrm{^1S_0}},M_I})}$ is the exact eigenstates for a Rabi-frequency $\Omega_{\mathrm{^1S_0}}$.
In \cref{fig:cooling_yb_b}, we calculate $\mathcal{F}$ as a function of the Rabi-frequency, and one obtains an ideal QND cooling scheme for sufficiently large Rabi-frequency.
Also, the Rabi-frequency one needs for $^{171}$ Yb is larger than the case of $^{87}$Sr and this is due to the larger Hyperfine interaction for $^{171}$Yb compared to $^{87}$Sr.

\onecolumngrid
\section{Tensor light shift }
\label{sec:tensor_light_shift}
In this section, we outline a detailed analysis of the energy light shift induced by coupling between the
 $\mathrm{5s5p^1P_1}$ to the  $\mathrm{5s15d^1D_2}\equiv c$. 
The light shift, with $\pi$ polarized light for state $\ket{M_J=0,M_I}$, given as,
\begin{equation}
    V_{\mathrm{LS}}^{(ac)}= \sum_{F',M_{F'}}\frac{1}{4\Delta_{F'}({\mathrm{^1D_2}})}\left\lvert\langle{c,F',M_{F'}}\lvert d_z\ket{a,M_J=0,M_I}\right\rvert^2.
    \label{eq:light_shift_Hamiltonian_d_app}
\end{equation}
where $\Delta_{F'}({\mathrm{^1D_2}})=\Delta_{\mathrm{^1D_2}}-\left[E_{F'}(c)-E_{M_J=0}(a)\right]=\Delta_{\mathrm{^1D_2}}+\delta_{F'}({\mathrm{^1D_2}})$.
Thus the light-shift interaction involves the coupling of states in $a$ where the good quantum numbers are the uncoupled basis $\ket{M_J,M_I}$ and for $c$ where the good quantum number is the coupled basis, $\ket{M_F}$.
However to find an electric dipole matrix element we need to work in either one of these bases for which one can use either the  decomposition,
\begin{equation}
    \ket{M_J,M_I}=\sum_F \braket{F,M_F}{I,M_I;J,M_J}\ket{F,M_F},
    \label{eq:expression_basis_uncoupled_to_coupled}
\end{equation}
or 
\begin{equation}
    \ket{F',M_{F'}}=\sum_{I',J}\braket{M_{I'},M_{J'}}{F,M_{F'}}\ket{M_{I'},M_{J'}}.
    \label{eq:expression_basis_coupled_to_uncoupled}
\end{equation}
Using \cref{eq:expression_basis_uncoupled_to_coupled}, the light shift interaction coupling $a \to c$ for a $\pi$ polarized light is given as,
\begin{equation}
    \begin{aligned}
V_{\mathrm{LS}}^{(ac)} \propto &\sum_{F',M_{F'}}\frac{\Omega_{ac}^2}{4\Delta_{F'}({\mathrm{^1D_2}})}\left.\lvert \sum_{F,M_F} \bra{c,F',M_{F'}}d_z \ket{a,F,M_F}\bra{F,M_F}\ket{I,M_I;J,M_J=0}\right.\rvert^2,
    \end{aligned}
\end{equation}
To find the matrix element for a particular value of $M_I$, one can use the Wigner-Eckart theorem. 
Notice that,
    \begin{equation}
    \begin{aligned}        \bra{c,F',M_{F'}}d_z\ket{a,J,M_J,I,M_I}&=\bra{F',M_{F'}}\ket{J,M_J,I,M_I}\bra{c,J',M_J;I',M_{I'}}d_z\ket{a,J,M_J,I,M_I}.
    \end{aligned}
\end{equation}
For  a $\pi$ polarized light using the reduced dipole matrix element one obtains,
\begin{equation}    \bra{c,J',M_J;I',M_{I'}}d_z\ket{a,J,M_J,I,M_I}=\langle{c,J'}\lvert\lvert d\rvert \rvert {a,J}\rangle\bra{J',M_{J'}}\ket{J,M_J;1,0}\delta_{M_I,M_{I'}}.
\end{equation}
Thus for $J=1,J'=2$ and $M_{J}=0$, we get,
\begin{equation}
    \begin{aligned}        \bra{c,F',M_{F'}}d_z\ket{a,J,M_J,I,M_I}&\propto\langle{c,J'}\lvert\lvert d\rvert \rvert {a,J}\rangle \langle{F',M_{F'}=M_I}\rvert{2,0;I=\frac{9}{2},M_I}\rangle\bra{2,0}\ket{1,0,1,0},
    \end{aligned}
\end{equation}
which in turn gives,
\begin{equation}
    V_{\mathrm{LS}}^{(ac)}(M_I)=\Omega_{ac}^2  \sum_{F'}\frac{1}{4\Delta_{F'}}\lvert\langle F',M_{F'}=M_I \rvert2,0; \frac{9}{2},M_I \rangle \rvert^2  \lvert\langle 2,0 \rvert \rangle 1,0;1,0\rvert^2 .
    \label{eq:light_shift_Hamiltonian_dd}
\end{equation}
where $\Omega_{ac}^2=E_{\mathrm{L}}^2  \langle{c,J'}\lvert\lvert d\rvert \rvert {a,J}\rangle^2$ and $E_{\mathrm{L}}^2$ is the proportionality constant.
For the case when we detune far away from all the hyperfine states i.e $\Delta_{\mathrm{^1D_2}}\gg \delta_{F'}$, we get,
\begin{equation}
    V_{\mathrm{LS}}^{(ac)}(M_I)=V_0^{ac}\left(1-\sum_{F'}\frac{\delta_{F'}}{\Delta_{\mathrm{^1D_2}}}\lvert\langle F',M_{F'}=M_I \rvert2,0; \frac{9}{2},M_I \rangle \rvert^2 \right)
\end{equation}
where $V_0^{ac}=\frac{\Omega_{ac}^2}{4}\lvert\langle 2,0  \rvert  1,0;1,0\rangle \rvert^2$.
In the other regime when we work closely detuned to $F'=13/2$, one can obtain,
\begin{equation}
     V_{\mathrm{LS}}^{(ac)}(M_I)\approx V_0^{ac}\lvert\langle F'=\frac{13}{2},M_{F'}=M_I \rvert2,0; \frac{9}{2},M_I \rangle \rvert^2.
\end{equation}
By empirically fitting,
\begin{equation}
    \lvert\langle F'=\frac{13}{2},M_{F'}=M_I \rvert2,0; \frac{9}{2},M_I \rangle \rvert^2\approx 0.3-0.017M_I^2+2.3 \cross10^{-4}M_I^4.
\end{equation}
The quadratic behavior is not familiar for a light shift (usually at most quadratic in nature). 
Here it arises from the way in which the nucleus is coupling to the electron through $J=2$ (which is quadrupolar rather than dipolar). 
This quadratic term indicates that one can cancel the energy light arising from the hyperfine perturbation in the state $a$ which also has a quadratic term from the perturbation theory analysis.

To further understand the light-shift Hamiltonian, one can use the following expansion,
\begin{small}
     \begin{equation}
      \begin{aligned}
    &\left\lvert \bra{F',M_{F'}}d_z\ket{J,M_J,I,M_I} \right\rvert^2=\left\lvert\sum_{F}\bra{F',M_{F'}}d_z\ket{F,M_F}\braket{F,M_F}{J=1,M_J=0,I=\frac{9}{2},M_I}\right\rvert^2,\\
    &=\sum_{F_1,F_2 } \bra{F_1,M_{F_1}}d_z\ket{F',M_{F'}}
    \bra{F',M_{F'}}d_z\ket{F_2,M_{F_2}}
    \braket{F_1,M_{F_1}}{J=1,M_J=0,I=\frac{9}{2},M_I} \braket{F_2,M_{F_2}}{J=1,M_J=0,I=\frac{9}{2},M_I},\\
    &=\sum_{F_1,F_2 } \bra{F_1,M_I}d_z\ket{F',M_I}
    \bra{F',M_I}d_z\ket{F_2,M_I}
    \braket{F_1,M_I}{J=1,M_J=0,I=\frac{9}{2},M_I} \braket{F_2,M_I}{J=1,M_J=0,I=\frac{9}{2},M_I}.\\
    \end{aligned}
\end{equation}  
\end{small}
Also one can write,
\begin{equation}
    \bra{F_1,M_I}d_z\ket{F',M_I}=\braket{F_1,M_I}{1,0,F',M_I} \langle F_1\lvert\lvert d_z\rvert \rvert F'\rangle,
\end{equation}
thus the light-shift interaction here is not restricted to a single angular momentum ground state $F$.
This is reflected in the fact that the light-shift does not have the familiar scalar-vector-tensor form in terms of the polynomials in $(F_x,F_y,F_z)$.
Working in a regime closely detuned to $c,F'=13/2$, one obtains,
\begin{equation}
     V_{\mathrm{LS}}^{(ac)}(M_I)\approx V_0^{ac}\left\lvert\langle F'=\frac{13}{2},M_{F'}=M_I \rvert2,0; \frac{9}{2},M_I \rangle \right \rvert^2,
\end{equation}
where $V_0^{ac}=\frac{\Omega_{ac}^2}{4}\lvert\langle 2,0  \rvert  1,0;1,0\rangle \rvert^2$.
By empirically fitting this as a function of $M_I$ we find,
\begin{equation}
\begin{aligned}
    &\left\lvert\langle F'=\frac{13}{2},M_{F'}M_I \rvert2,0; \frac{9}{2},M_I \rangle \right\rvert^2\\
    &\approx 0.3-0.017M_I^2+2.3 \cross10^{-4}M_I^4.
\end{aligned}
        \label{eq:empirical_fitting}
\end{equation}
The quartic behavior is not familiar for a light shift (usually at most quadratic in nature). 
Here it arises from how the nucleus is coupling to the electron through $J=2$ (which is quadrupolar rather than dipolar). 
The dominant quadratic term can be used to cancel the energy light arising from the hyperfine perturbation in the state $a$, which also has a quadratic term from the perturbation theory analysis.

To further understand the quartic behavior in the tensor light shift, one can expand the $\ket{\mathrm{5s5p^1P_1}, M_J=0}$ in the coupled basis.
Working close to resonance for $F'=13/2$  and using the fact that the  dipole allowed interaction only allows  $F=F'\pm 1$ the only matrix element we need to consider is,
    \begin{equation}
    \begin{aligned}
        &\bra{c,F'=13/2,M_I}d_z\ket{a,1,0;9/2,m_I}        =\bra{c,F'=13/2,M_I}d_z\ket{a,F=11/2,M_I}  \braket{F=11/2,M_I}{1,0;9/2,M_I},\\
        &= \rme{c,F'=13/2}{d_z}{a,F=11/2}\braket{F'=13/2,M_I}{1,0;F=11/2,M_I}\braket{F=11/2,M_I}{1,0;9/2,M_I},\\
        &=\mathcal{O}^{J',F}_{J,F}\rme{J'=2}{d_z}{J=1}\braket{F'=13/2,M_I}{1,0;F=11/2,M_I}\braket{F=11/2,M_I}{1,0;9/2,M_I},\\
        &=\rme{J'=2}{d_z}{J=1}\braket{F'=13/2,M_I}{1,0;F=11/2,M_I}\braket{F=11/2,M_I}{1,0;9/2,M_I},
    \end{aligned}
    \label{eq:knowing_the_quartic_dependence_2_app}
\end{equation}
where $\mathcal{O}^{J',F'}_{J,F}$ is the relative oscillator strength defined as,

    \begin{equation}
   \mathcal{O}^{J',F'}_{J,F}=(-1)^{F'+1+F+I}\sqrt{(2J'+1)(2F+1)}\begin{Bmatrix}
       F' & I & J'\\
       J & 1 &F
   \end{Bmatrix}, 
\end{equation}

\noindent and for the case of $J'=2,J=1,F'=13/2,F=11/2$, we get $\mathcal{O}^{J',F'}_{J,F}=1$.
Using the following  property of Clebsch-Gordan coefficients \cite{ME_Rose_angular_momentum},
\begin{equation}
    \braket{j+1,m}{1,0;j,m}=\sqrt{\frac{(j+1)^2-m^2}{(2j+1)(j+1)}},
\end{equation}
 gives,
\begin{equation}
    \begin{aligned}
        &\braket{F'=13/2,M_I}{1,0;F=11/2,M_I}=\frac{1}{2}\sqrt{\frac{169-4M_I^2}{78}}\\
        &\braket{F=11/2,M_I}{1,0;I=9/2,M_I}=\frac{1}{2}\sqrt{\frac{121-4M_I^2}{55}}.
    \end{aligned}
\end{equation}
Thus we get the contribution of the tensor light-shift interaction for the case of close to resonance to $F'=13/2$ as,
\begin{equation}
\begin{aligned}
   V_{\mathrm{LS}}^{ac}= &\lvert \bra{c,F',M_{F'}} d_z \ket{a,J,M_J;I,M_I} \rvert^2\\
    &=V_0^{ac}\left(0.298-0.0169M_I^2+0.000233M_I^4\right).
    \end{aligned}
\end{equation}
This is approximately the same expression we got from the empirical fitting in \cref{eq:empirical_fitting}.


\bibliography{reference}
\end{document}